\documentclass[]{aa}
\usepackage{multirow}
\usepackage{gensymb}
\usepackage{subfigure}
\usepackage{graphicx}
\usepackage{natbib}
\usepackage{appendix}
\usepackage{txfonts}
\titlerunning{Chemical composition of the circumstellar disk around AB Aurigae}
\authorrunning{Pacheco-V\'azquez, S. et al. 2015}

\begin{document} 
\newcommand\Msun{M$_{{\odot}}$\:}
\newcommand\pow[1]{10$^{#1}$}
\newcommand\Rin{R$_{\mathrm{in}}$\:}
\newcommand\Rout{R$_{\mathrm{out}}$\:}
\newcommand\kms{km~s$^{-1}$}
\newcommand\Lsun{L$_{{\odot}}$}
   
\title{Chemical composition of the circumstellar disk around AB Aurigae}

   \subtitle{}

   \author{ S. Pacheco-V\'azquez\inst{1},
	    A. Fuente\inst{1},
	    M. Ag\'undez \inst{2},
	    C. Pinte \inst{6,7},
	    T. Alonso-Albi\inst{1},
	    R. Neri\inst{3},
            J. Cernicharo\inst{2},
	    J. R. Goicoechea\inst{2},
	    O. Bern\'e\inst{4,5},
	    L. Wiesenfeld\inst{6},
            R. Bachiller\inst{1},
            \and
            B. Lefloch\inst{6}
%\thanks{Just to show}
          }

   \institute{
     Observatorio Astron\'omico Nacional (OAN), Apdo 112, E-28803 Alcal\'a de Henares, Madrid, Spain\\
          \email{s.pacheco@oan.es, a.fuente@oan.es}\and
     Instituto de Ciencia de Materiales de Madrid, ICMM-CSIC, C/ Sor Juana Inés de la Cruz 3, E-28049 
     Cantoblanco, Spain\\
          \email{marcelino.agundez@icmm.csic.es}\and
     Institut de Radioastronomie Millim\'etrique, 300 Rue de la Piscine, F-38406 Saint Martin d'Hères, 
     France\\\and
     Université de Toulouse, UPS-OMP, IRAP, Toulouse, France\and
     CNRS, IRAP, 9 Av. colonel Roche, BP 44346, F-31028 Toulouse cedex 4, France\\\and
     Institut de Plan\'etologie et d'Astrophysique de Grenoble (IPAG) UMR 5274, Université UJF-Grenoble 
     1/CNRS-INSU, F-38041 Grenoble, France\\\and
     UMI-FCA, CNRS/INSU, France (UMI 3386), and Dept. de Astronom\'ia, Universidad de Chile, Santiago, 
     Chile\\
          \email{christophe.pinte@obs.ujf-grenoble.fr}
             }

   \date{Received September 15, 1996; accepted March 16, 1997}

%%%%%%%%%%%%%%%%%%%%%%%%%%%%%%%%%%%%%%%%%%%%%%%%%%%%%%%%%%%
%		               )\._.,--....,'``.      	%%%
%%	 .b--.    	      /;   _.. \   _\  (`._ ,.   %%
%%%	`=,-,-'~~~   	     `--- (,_..'--(,_..'`-.;.'    % 
%%%%%%%%%%%%%%%%%%%%%%%%%%%%%%%%%%%%%%%%%%%%%%%%%%%%%%%%%%%
\abstract
   {}
   {Our goal is to determine the molecular composition of the circumstellar disk
around AB Aurigae (hereafter, AB Aur). AB Aur is a prototypical Herbig Ae star
and the understanding of its disk chemistry is paramount for understanding the chemical evolution of the gas in warm disks. 
}
   {We used the IRAM 30-m telescope to perform a sensitive search for molecular
lines in AB Aur as part of the IRAM 
Large program ASAI (A Chemical Survey of Sun-like Star-forming Regions). These
data were complemented with interferometric
observations of the HCO$^+$ 1$\rightarrow$0 and C$^{17}$O 1$\rightarrow$0 lines
using the IRAM Plateau de Bure Interferometer (PdBI). 
Single-dish and interferometric data were used to constrain chemical models.}
   {Throughout the survey, several lines of CO and its isotopologues, HCO$^+$,
H$_2$CO, HCN, CN, and CS, were detected.  
In addition, we detected the SO 5$_4$$\rightarrow$3$_3$ and
5$_6$$\rightarrow$4$_5$ lines, confirming the previously tentative detection.
Compared to other T Tauri and Herbig Ae disks, AB Aur presents low HCN
3$\rightarrow$2/HCO$^+$ 3$\rightarrow$2 and CN 2$\rightarrow$1/HCN
3$\rightarrow$2 line intensity ratios, similar to other transition disks.
AB Aur is the only protoplanetary disk detected in SO thus far, and its
detection is consistent with interpretation of this disk being younger than
those associated with T Tauri stars.
}
   {We modeled the line profiles using a chemical model and a radiative transfer
3D code. Our model assumes a flared disk in hydrostatic equilibrium. 
The best agreement with observations was obtained for a disk with a mass of 0.01
M$_\odot$, \Rin=110 AU, \Rout=550 AU, a surface density radial index of 1.5, and
an
inclination of 27$\degree$. The intensities and line profiles were reproduced
within 
a factor of $\sim$2 for most lines. This agreement is reasonable considering
the simplicity of our model that neglects any structure within the disk.
However, 
the HCN 3$\rightarrow$2 and CN 2$\rightarrow$1 line intensities were predicted to be more 
intense by a factor of $>$10. We discuss several scenarios to explain this discrepancy.}
    \maketitle

%%%%%%%%%%%%%%%%%%%%%%%%%%%%%%%%%%%%%%%%%%%%%%%%%%%%%%%%%%%
%		               )\._.,--....,'``.      	%%%
%%	 .b--.    	      /;   _.. \   _\  (`._ ,.   %%
%%%	`=,-,-'~~~   	     `--- (,_..'--(,_..'`-.;.'    % 
%%%%%%%%%%%%%%%%%%%%%%%%%%%%%%%%%%%%%%%%%%%%%%%%%%%%%%%%%%%
\section{Introduction}

%Los discos son...
Circumstellar disks are commonly observed around pre-main sequence stars (e.g.,
\citealp{Howard2013, Strom1989}). 
The formation of disks, together with ejecta phenomena such as outflows and
jets, dissipate away the excess of angular momentum that prevents accretion from
the parent cloud. The chemical composition of dust and gas contained in these
disks provides information about the initial conditions in the formation of
planetary systems \citep{Dutrey2014}.

%La quimica de los discos
The comprehension of chemistry in disks is an important step toward understanding of the formation of complex organic, even prebiotic molecules on
planets. However, the disk chemistry is a very unexplored field from the
observational point of view with very few molecular detections. 
This scarcity of molecules seems more accentuated in disks around Herbig Ae
stars \citep{Oberg2011}. This is mainly due to the low molecular abundances in a
gas disk that itself has a low mass content. The ultraviolet radiation 
from the central star photodissociates molecules in the surface layers of the
disk. Deeper in the midplane, the temperatures drop, and all the detectable
molecules freeze out onto dust grains. As a result, molecules can only survive in the
gas phase inside a thin layer.
For F and A stars with effective temperatures in the range between 6000 to 10000
K, the UV-photons penetrate deeper into the disk than the colder M and K stars
(T$_{\rm eff}$ $\sim$ 2500 - 5000K), causing the drop of the molecular detection
rates.
Most species detected are simple molecules, molecular radicals, and ions, such as
CO, $^{13}$CO, C$^{18}$O, CN, CS, C$^{34}$S, C$_2$H, HCN, H$^{13}$CN, HNC, DCN,
HC$_3$N, HCO$^+$, H$^{13}$CO$^+$, DCO$^+$, H$_2$D$^+$, N$_2$H$^+$, c-C$_3$H$_2$,
H$_2$CO, H$_2$O, and HD (e.g., \citealp{Kastner1997, Dishoeck2003, Thi2004,
Qi2008, Guilloteau2006, Pietu2007, Dutrey2007}).

%Los modelos
Unfortunately, most disks remain unresolved even with the largest millimeter
interferometers. 
A detailed study of the chemical composition of the gas in a disk requires not
only high angular resolution observations in dust continuum and molecular lines,
but also accurate chemical and physical models.
These models help to constrain the disk structure in accordance with the
observations and calculate the molecular abundance profiles. In recent years,
the chemical and radiative transfer models have improved their performances,
(see, e.g., \citealp{Thi2013,Pinte2010,Nomura2009,Agundez2008,Dutrey2007}).

%AB Aur tiene un disco...
Herbig Ae/Be stars are intermediate-mass pre-main sequence sources that emit
much stronger thermal UV radiation than do T Tauri stars (TTs). Therefore, their
circumstellar disks are warmer and more ionized.
Our target, AB Aurigae (hereafter, AB Aur) is one of the best-studied Herbig Ae
stars that host a prototypical Herbig Ae disk. It has a spectral type A0-A1
\citep{Hernandez2004}.
It has a M$_\star$$\sim$ 2.4 \Msun, a T$_{\rm eff}$ $\sim$ 9500 K, and it is
located at a distance of 145 pc \citep{Ancker1998}. 
The disk around AB Aur shows a complex structure. It is larger (R$_{\rm
out}$$\sim$1100 AU, $\approx$7$\arcsec$), than those around TTs, and it shows spiral-arm features traced by millimeter continuum emission, at about 140 AU from the
star \citep{Pietu2005}. 

By modeling the $^{12}$CO and its isotopologue lines obtained from subarcsec
imaging with the Plateau de Bure Interferometer (PdBI), it was found that,
contrary to typical disks associated with TTs, the AB Aur disk is warm ($>$25 K
all across the disk) and shows no evidence of CO depletion \citep{Pietu2005}.
\citet{Schreyer2008} carried out a chemical study of the disk around AB Aur with
the PdBI searching for the HCO$^+$ 1$\rightarrow$0, CS 2$\rightarrow$1, HCN
1$\rightarrow$0, and C$_2$H 1$\rightarrow$0
lines, but with only one detection, the HCO$^+$ 1$\rightarrow$0 line. They propose that
the poor molecular content of this disk is because the UV-photons dissociate
molecules.
More recently, \citet{Fuente2010} have carried out a molecular search using the IRAM
30-m telescope. 
As a result, they detected the SO 3$_{4}$$\rightarrow$2$_{3}$ line. SO had
never been detected before in a T Tauri or 
Herbig Ae disk. Its detection provided further support to the interpretation of
a warmer chemistry in this kind of Herbig Ae disk, even if there are some
observations that suggest the possibility that the SO emission could be coming
from an outflow or from an envelope rather than from the circumstellar disk
\citep{Guilloteau2013}.

%En este trabajo
In this paper we investigate the molecular composition on the AB Aur disk from data
obtained as part of the IRAM Large program ASAI (A Chemical Survey of Sun-like
Star-forming Regions) (PIs: R. Bachiller, B. Lefloch). These data were
complemented with interferometric observations using the IRAM PdBI. 
Single-dish and interferometric data were used to constrain chemical models.
A spectral survey is a powerful tool for characterizing the chemical composition of
a source. It is the only way to get a complete census of all molecular species
and to provide several lines of the same molecule, giving the possibility of
multiline analysis and modeling.

To reach our aim, we use an updated version of the chemical model described in
\citet{Agundez2008} and the 3D radiative transfer code MCFOST (see, e.g.,
\citealp{Pinte2006, Pinte2009}).

\begin{figure*}[ht!]
   \centering
   \includegraphics[scale=1.05]{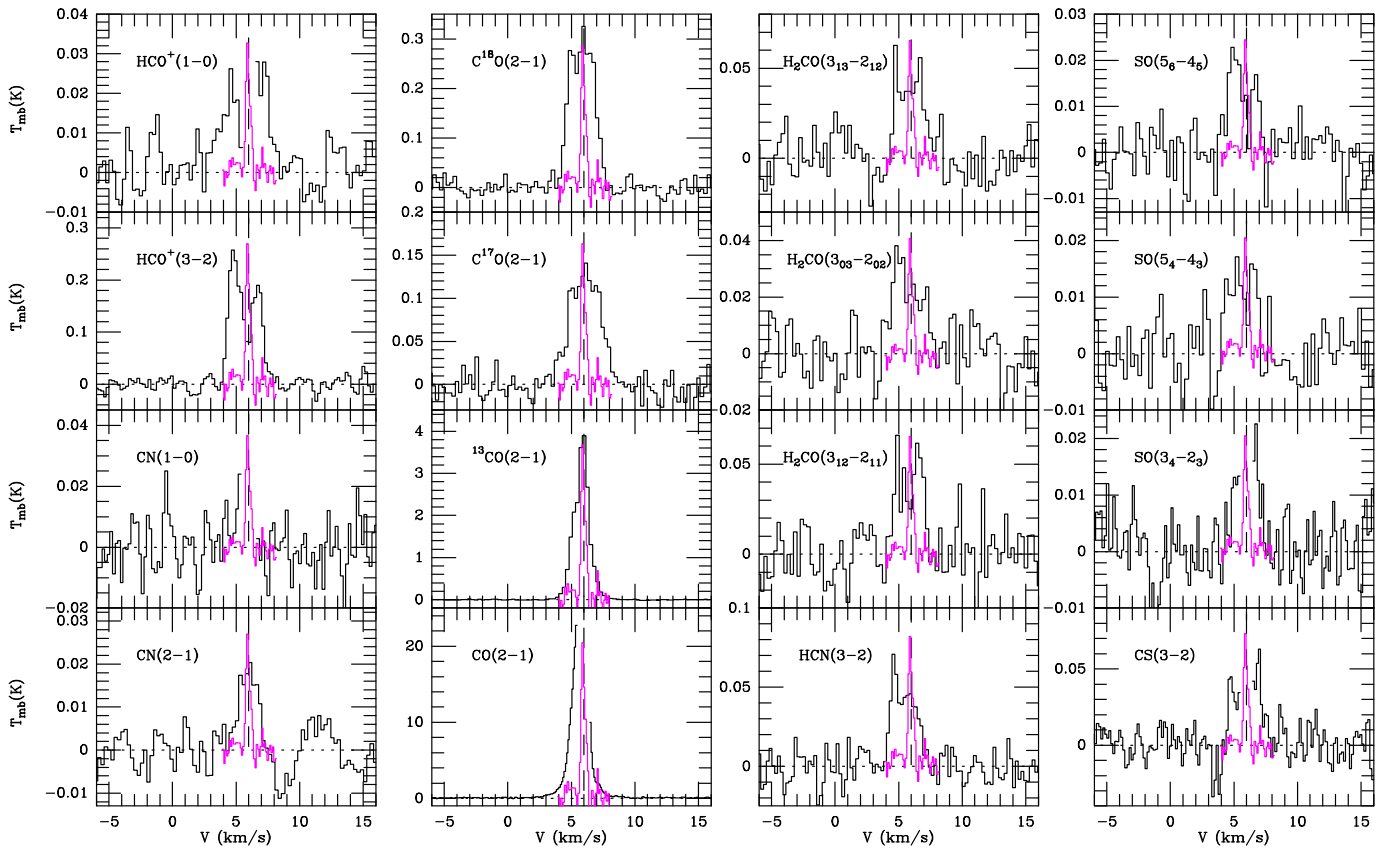}
   \caption{Spectral lines detected toward AB Aur with the IRAM 30-m telescope.
The transitions CO 2$\rightarrow$1, HCO$^{+}$ 1$\rightarrow$0, SO
3$_4$$\rightarrow$2$_3$, CN 1$\rightarrow$0, and CS 3$\rightarrow$2 were
taken from \cite{Fuente2010}, where cloud velocities were blanked. 
The HCN 3$\rightarrow$2 is an averaged spectrum from previous observations by
Fuente et al. (2010) and ASAI data. Magenta line shows the narrow C$^{18}$O
1$\rightarrow$0 line profile that traces the emission from the foreground cloud,
and for visualization purposes the C$^{18}$O 1$\rightarrow$0 line (T$_{\rm
peak}$$\sim$ 0.7 K) is scaled to fit
in each panel.}
   \label{Fig:detecciones}
\end{figure*}

%%%%%%%%%%%%%%%%%%%%%%%%%%%%%%%%%%%%%%%%%%%%%%%%%%%%%%%%%%%
%		               )\._.,--....,'``.      	%%%
%%	 .b--.    	      /;   _.. \   _\  (`._ ,.   %%
%%%	`=,-,-'~~~   	     `--- (,_..'--(,_..'`-.;.'    % 
%%%%%%%%%%%%%%%%%%%%%%%%%%%%%%%%%%%%%%%%%%%%%%%%%%%%%%%%%%%
\section{Observations}
\label{Observations}

\subsection{IRAM 30-m radiotelescope}
We carried out a spectral survey toward AB Aur 
($\alpha_{\rm J2000}$=04$\mathrm{h}$\,55\,$\mathrm{m}$ 45.8\,$\mathrm{s}$,\,
$\delta_{\rm J2000}$=30$^{\circ}$ 33' 04''.2) 
as part of the IRAM Large program ASAI using the IRAM 30-m telescope at Pico
Veleta (Granada, Spain).
Several observing periods were scheduled on July and February 2013 and  
January and Mars 2014, to cover the 1, 2, and 3 mm bands shown in Table
\ref{table:bands}. 
The Eight MIxer Receivers (EMIR) and the fast Fourier Transform Spectrometers
(FTS) with a spectral resolution of 200\,kHz
were used for the whole project.
The observing procedure was wobbler switching with a throw of 200$\arcsec$ to
ensure flat baselines and to avoid 
possible contamination from the envelope toward this young disk. In
Table 1 we show the beam efficiency, half power beam width (HPBW), spectral
resolution, and sensitivity achieved in each observed frequency band. 

The data reduction and the line identification were carried out with the package
CLASS of GILDAS software \citep{Maret2011}. Three databases were used to
identify the 
lines: (1) Cologne Database for Molecular Spectroscopy (CDMS; see
\citealp{Muller2005}), (2) the molecular spectroscopy database of Jet Propulsion 
Laboratory (JPL; see \citealp{Pickett1998}), and (3) MADEX \citep{Cernicharo2012}.

\begin{table}
\caption{Observed band ranges with IRAM 30-m telescope.}           
\label{table:bands}      
\centering   
\small                      
\begin{tabular}{p{0.7cm}p{1.8cm}p{0.6cm}p{1cm}p{0.9cm}p{1.1cm}}       
\hline\hline     
Band	&Frequency	& rms	&$\Delta$v	&HPBW	&$\eta_b$\\
	& [GHz]		& (mK)	& (\kms)	&(\arcsec)&\\\hline
E090	&  84.5 - 96.3	&5-7	& 0.6  		& 29-25&0.85\\
E150	&  133.8- 144.8 &5-8	& 0.4		& 18-17&0.79\\
E230	&  200.5- 272.0	&5-16	& 0.3		& 12-9&0.67-0.6\\
           \hline                                   
\end{tabular}
\end{table}

\subsection{Plateau de Bure interferometer}

The interferometric observations were carried out in the second half of 2011
using the C and D configurations of the PdBI with six antennas.
This configuration provided a beam of
$\approx$5.41$\arcsec$$\times$4.31$\arcsec$ with a position angle (PA) of
129$\degree$ at the frequency of HCO$^+$ 1$\rightarrow$0 (89.1885~GHz) 
and of $\approx$5.41$\arcsec$$\times$4.31$\arcsec$ with a PA of 139$\degree$ at
the frequency of C$^{17}$O 1$\rightarrow$0 (112.3593~GHz). 
During the observations, one 20 MHz bandwidth correlator unit was placed at the
frequency of the HCO$^+$ 1$\rightarrow$0 
line, providing a spectral resolution of $\sim$39 kHz (= 0.13 \kms). 
The C$^{17}$O line was observed with the WideX correlator, which provides a
spectral resolution of $\sim$2~MHz (= 5.4 \kms). We used 3C454.3, MWC349,
3C84, 0552+398, and 0415+379 as phase 
and flux calibrators.

Data reduction and image synthesis were carried out using the GILDAS software. 
The channels free of line emission  were used to estimate the 
continuum flux that was subtracted from the spectral maps. To improve
the S/N, the 
HCO$^+$ image was created with a velocity 
resolution of 0.4~\kms.~The rms of the resulting cube was $\sim$6~mJy/beam. In
the case of C$^{17}$O, we kept the original spectral resolution
and produced a cube with rms of $\sim$2~mJy/beam.

\begin{figure}[b!]
    \centering
    \includegraphics[scale=0.85]{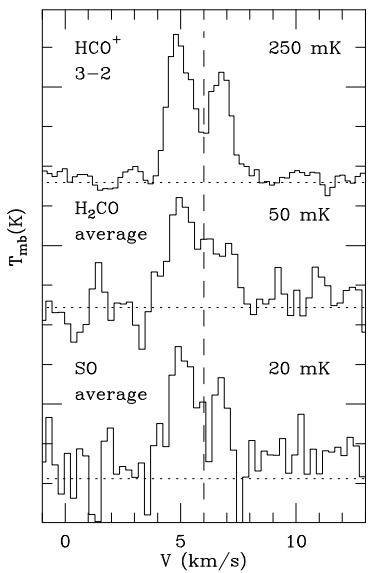}
    \caption{Comparison of the averaged spectra of the SO and H$_2$CO lines with
the HCO$^+$3$\rightarrow$2 line profile. In the \emph{top} of these panels we
have the HCO$^+$3$\rightarrow$2 profile showing the double-peak shape that is
characteristic of a rotating disk. In the \emph{middle} we can see the averaged
H$_2$CO profile that keep the same double-peak feature in the same range of
velocities. In the \emph{bottom} the averaged SO profile is in complete
agreement with the two other spectra.}
    \label{Fig:comp_averaged}%
\end{figure}
%%%%%%%%%%%%%%%%%%%%%%%%%%%%%%%%%%%%%%%%%%%%%%%%%%%%%%%%%%%
%		               )\._.,--....,'``.      	%%%
%%	 .b--.    	      /;   _.. \   _\  (`._ ,.   %%
%%%	`=,-,-'~~~   	     `--- (,_..'--(,_..'`-.;.'    % 
%%%%%%%%%%%%%%%%%%%%%%%%%%%%%%%%%%%%%%%%%%%%%%%%%%%%%%%%%%%
\section{Results}
\label{Results}
\subsection{IRAM 30-m}

In Table \ref{table:detection_list}, we show the list of molecular lines
detected at a level $>$5$\sigma$. This high S/N is required for a secure
detection and for having good line profiles for our analysis.
For completeness, we also include those lines previously detected by Fuente et
al. (2010).

The spectra are shown in Fig. \ref{Fig:detecciones}.
The lines of CO and $^{13}$CO show the narrow line feature
($\Delta$v $\sim$ 0.5 \kms) typical of the lines coming from the foreground
cloud (see discussion in Fuente et al. 2010). This suggests that their
integrated line emission is dominated by the cloud emission. However, all the
other lines show the double peak profile characteristic of rotating disks.
In Sect. \ref{Comp_IRAM-30m-PdBI} we compare the flux of the HCO$^+$
1$\rightarrow$0 line listed in Table \ref{table:detection_list}
with the one obtained from our interferometric map and demonstrate that the HCO$^+$
1$\rightarrow$0 emission is coming from the disk. For this reason, we
adopt HCO$^+$ as a pattern for the disk emission.

The AB Aur disk presents some peculiarities. In a first eye verification of the
detected lines, we realized that HCN 3$\rightarrow$2 line in AB Aur has an
intensity comparable to the less common species like H$_2$CO. \"{O}berg et al. (2010,
2011) carried out a disk imaging survey with SMA toward twelve sources that cover
a wide range of stellar luminosities among the TT and Herbig Ae stars. They
show that HCN is strong in disks: HCN 3$\rightarrow$2 and CN 2$\rightarrow$1
are the strongest lines after HCO$^+$ and CO, and are twice as intense than
the H$_2$CO at 1mm.

Besides this, SO has never been detected in a protoplanetary disk. It is, however,
very abundant in the shocks produced by bipolar outflows, (see, e.g.,
\citealp{Bachiller1997}). In AB Aur disk, we have detected two transitions of
SO, 5$_4$$\rightarrow$4$_3$, and 5$_6$$\rightarrow$4$_5$ in the 1mm band, and
Fuente et al. (2010) detected the SO 3$_4$$\rightarrow$2$_3$ at 2mm.
All the SO lines present double-peak line profiles consistent with the
interpretation of the emission coming from the circumstellar disk of AB Aur (see
Fig. \ref{Fig:detecciones}). 

In Fig. \ref{Fig:comp_averaged} we present a comparison between the averaged
spectra of H$_2$CO and SO 1mm lines with the line profile of HCO$^+$
3$\rightarrow$2. We
only averaged the lines observed in the 1mm band with similar HPBW, which makes
the comparison more reliable.
The averaged spectra of both H$_2$CO and SO match the double-peak shape in the
same range of velocities as in the case of HCO$^+$ line profile. Waiting for
further confirmation with interferometric observations, we consider that the
circumstellar disk is the most likely origin for the SO emission.

\begin{table}[]
	\caption{Detections in AB Aur disk.}           
	\label{table:detection_list}      
	%\centering                      
	\begin{tabular}{p{1cm} p{1.3cm} l l p{1cm} p{0.8cm}}       
	\hline\hline
Molecule  & Transition    & Frequency  & E$_{\mathrm{up}}$&Area$^\star$&rms$^{\star\star}$\\
	   &              &    [GHz]   &    [K]     &{\small[mK\kms]}&[mK]\\\hline
$^{12}$CO & 2$\rightarrow$1  \:$\diamond$  & 230.538   & 16.6 &24388 &35\\
$^{13}$CO & 2$\rightarrow$1 	 	& 220.399   & 15.9 	&3 375 &5\\
C$^{17}$O & 2$\rightarrow$1		& 224.714   & 16.2 	&364 &8	\\
C$^{18}$O & 2$\rightarrow$1 	  	& 219.560   & 15.8 	&735 &9	\\
HCO$^{+}$ & 1$\rightarrow$0 \:$\diamond$& 89.188    & 4.3 	&47 &4	\\
	  & 3$\rightarrow$2 	  	& 267.557   & 25.7 	&610 &15\\
H$_2$CO   & 3$_{03}$$\rightarrow$2$_{02}$ & 218.222   & 21 	&75 &7	\\
          & 3$_{13}$$\rightarrow$2$_{12}$ & 211.211   & 32.1 	&100 &10\\
          & 3$_{12}$$\rightarrow$2$_{11}$ & 225.697   & 33.5 	&114 &14\\
SO        & 5$_4$$\rightarrow$4$_3$   	& 206.176   & 38.6 	&29 &5	\\
          & 5$_6$$\rightarrow$4$_5$   	& 219.949   & 35 	&35 &6	\\
          & 3$_4$$\rightarrow$2$_3$   \:$\diamond$& 138.178   & 15.9 &26 &4\\
HCN	  & 3$\rightarrow$2		& 265.886   & 25.5	&144 &14\\
CN	  & 1$\rightarrow$0 \:$\diamond$	& 113.490   & 5.5&26 &6	\\
CN	  & 2$\rightarrow$1		& 226.874   & 16.3	&63 &8	\\
CS 	  & 3$\rightarrow$2\:$\diamond$	& 146.969   & 14.1	&75 &7	\\
	\hline                                   
\end{tabular}

\noindent
\raggedright
{\footnotesize
$\diamond$ Transitions previously detected by \citet{Fuente2010}. $^\star$
Integrated intensity area in the velocity range [3,8] \kms. $^{\star\star}$ rms
in a channel of  \,0.26 \kms~ in the 1mm band and 0.65 \kms~ in the 3mm band.}
\end{table}

%Non-detections
In Table \ref{table:non-detections}, we show the 3$\sigma$ upper limits to the column densities 
of some interesting species. The column densities were calculated with MADEX (Cernicharo, 2012)
assuming a line width of $\Delta v$ $\sim$4 \kms and a source size of $\sim$14 
arcsec (2000 AU at 145 pc). 
We adopted a molecular hydrogen density of 3.5$\times$\pow{6}cm$^{-3}$ for two representatives temperatures: 30 and 50~K.
The first one is similar to the value derived from the 
$^{13}$CO observations at 200 AU by Pi\'etu et al. (2005) and corresponds to the temperature 
in the midplane predicted by our model (see Sect. \ref{Modelization}). The value of 50 K is the mass-averaged 
kinetic temperature derived from our model.
We searched for the abundant species HNC, DCN, C$_2$H, DCO$^+$, N$_2$H$^+$,
HC$_3$N, and C$_3$H$_2$ that have been detected in other disks
\citep{Dutrey2014}. However, we did not detect any of these species. They have weak lines whose emission could be below our sensitivity limit. The
tentative detection of H$_2$S by \citet{Fuente2010} was not confirmed. We also
searched for HCS$^+$, HCO, SiO, and CH$_3$OH lines, which are abundant species in
young protostars. They have not been detected coming from any disk or around AB Aur.

\subsection{Plateau de Bure}

\begin{figure*}
	\centering
	\subfigure{\includegraphics[scale=0.5]{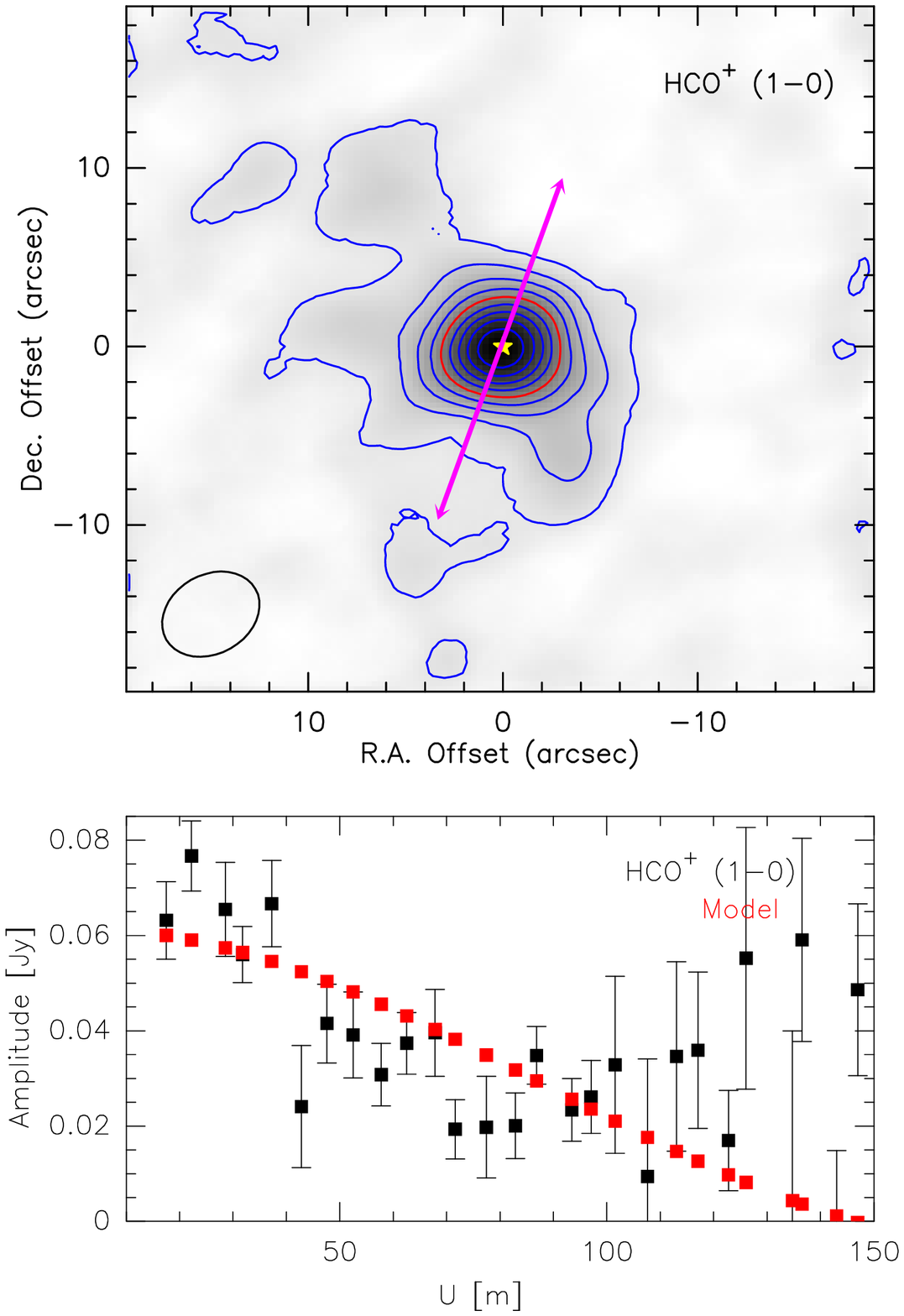}\label{Fig:map-model-HCOp}}
	\subfigure{\includegraphics[scale=0.5]{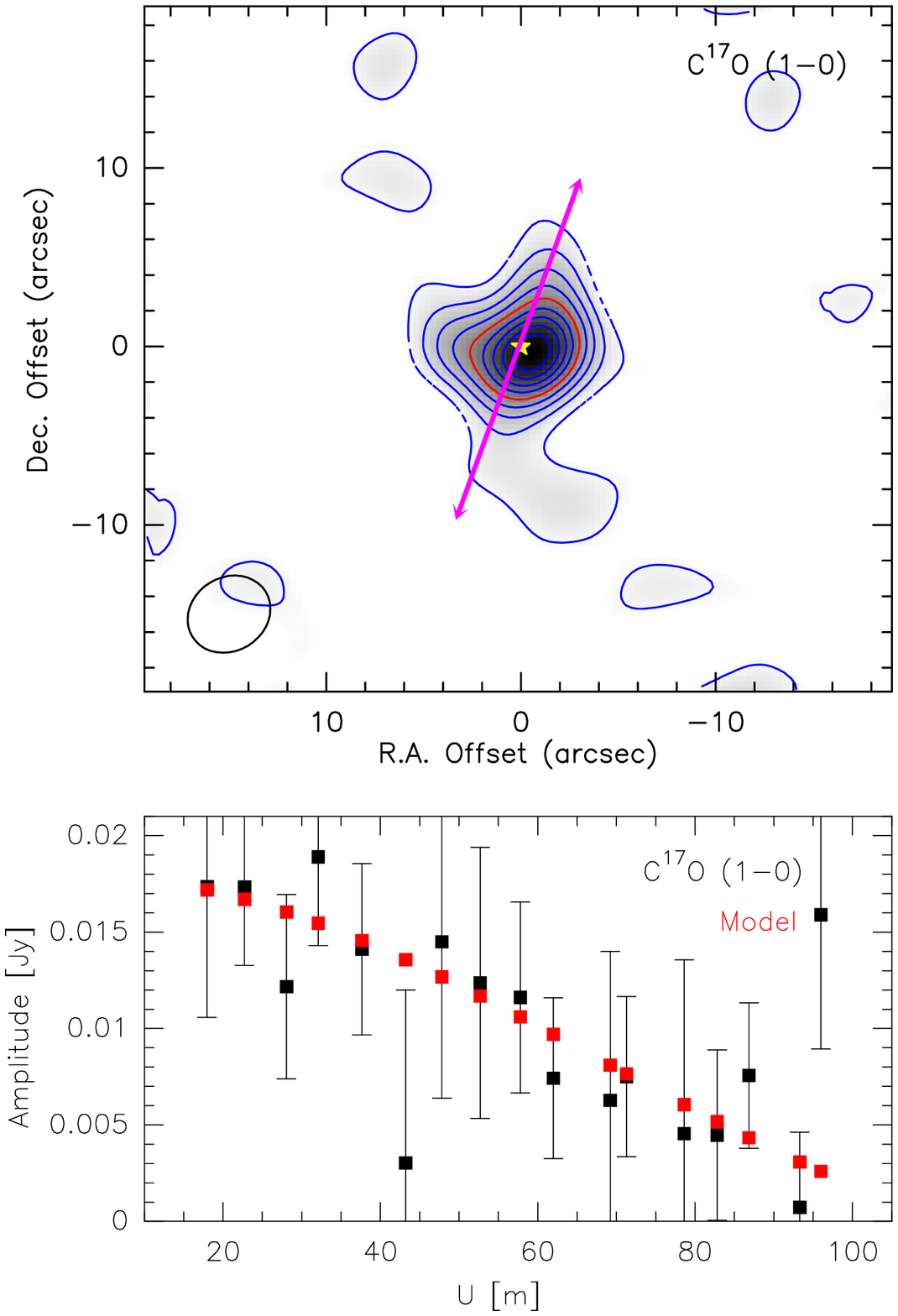}\label{Fig:map-model-C17O}}
	\caption{(\emph{Upper panels}) HCO$^+$ 1$\rightarrow$0 (left) and
C$^{17}$O 1$\rightarrow$0 (right) integrated velocity maps obtained with the
PdBI from 4 to 8 \kms. Blue contours levels are percentages of the maximum (0.15
Jy beam$^{-1}$ for C$^{17}$O and 0.22 Jy beam$^{-1}$ for HCO$^+$) from 10 to 90
in steps of 10. The position of the central source is marked by a yellow star.
The magenta arrow indicates the direction of the jet ($\sim$70$^\circ$) reported
by \cite{Rodriguez2014}. The red contour represents the half maximum power
contour. The ellipse in the left bottom corner indicates the synthesized beam
sizes: 5.41$\arcsec$ x 4.31$\arcsec$ with a position angle (PA) of 39$^\circ$
for HCO$^+$ 1$\rightarrow$0 and 4.58$\arcsec$ x 3.98$\arcsec$ with a PA of
49$^\circ$ for the C$^{17}$O. (\emph{Lower panels})
	Corresponding \emph{uv} plots to the upper maps. The \emph{uv} plot
shows the real part of visibilities as a function of the \emph{uv} distance,
binned in circular rings in steps of 5m.}
	\label{Fig:maps-models}
\end{figure*}

The upper panels of Fig. \ref{Fig:maps-models} show the contours of the
integrated-velocity intensity maps of HCO$^+$ 1$\rightarrow$0 (on the left) and
C$^{17}$O 1$\rightarrow$0 (on the right). Both maps are integrated in the
velocity range between [4,8] \kms. 
The emission coming from the disk traced by HCO$^+$ 1$\rightarrow$0
line is elongated and almost perpendicular to the emission traced by the
collimated outflow recently reported by \citet{Rodriguez2014} (direction
angle$\sim$70\degree). 
Moreover, the rotation axis of HCO$^+$ is consistent with that of the CO gas
disk traced by \citet{Pietu2005} ($\sim$50\degree).

The bottom panels of Fig. \ref{Fig:maps-models} show the real part of the
visibilities of the AB Aur molecular emission as a function of the \emph{uv}
radius. Each black point is an azimuthal average of the visibilities inside
circular rings in steps of 5 m from 0 to 200 m. For the \emph{uv} analysis, we
tried to fit different models, but the data quality does not allow us to
distinguish between circular and ring disk models. Red points indicate the best
fit to the \emph{uv} plot, using a circular disk model. 
In the case of the \emph{uv} plot of the C$^{17}$O emission, we note that it is
well fitted by a circular disk model centered on $\alpha_{\rm
J2000}$=04$\mathrm{h}$\,55\,$\mathrm{m}$ 45.81\,$\mathrm{s}$,\, $\delta_{\rm
J2000}$=30$^{\circ}$ 33' 03''.83, with a flux density of 18$\pm$3 mJy and
6$\pm$0.8\arcsec of diameter. 
However, for the case of the HCO$^+$, the fit is not good at long baselines.
Although there is much noise in the HCO$^+$ data, the \emph{uv} plot suggests
that there might be two components, a compact unresolved component, which is
responsible for the flattening of the \emph{uv} visibility amplitude at large (>100 m)
baselines, and an extended component (resolved) with a decreasing trend of the
\emph{uv} visibility amplitudes with radius at small \emph{uv} radii. The flux
density for this HCO$^+$ model was 62$\pm$3 mJy, and it had the same diameter of
5.8$\pm$0.3\arcsec. The fluctuations observed at large \emph{uv} radii indicate
that the morphology of the source is complex, consistently with the complexity
of the structures revealed by previous observations of the region around AB Aur
(e.g., \citealp{Pietu2005, Schreyer2008, Rodriguez2014}).

%C17O
% C_DISK   Flux        =     0.01802 (  0.00261)
% C_DISK   Diam.       =     5.95934 (  0.79413)
%
%HCO+
% C_DISK   Flux        =     0.06162 (  0.00327)
% C_DISK   Diam.       =     5.78181 (  0.34709)

\begin{table}[h!]
	\caption{Non-detections in AB Aur disk.$^1$}           
	\label{table:non-detections}      
	\centering           
	\small              
	\begin{tabular}{p{1cm} p{1cm} p{.45cm} p{.7cm} p{.65cm} p{.85cm} c}       
	\hline\hline   
Species		&Frequency	&rms&	$\delta$v& 3$\sigma$&N(50K)&N(30K)\\
		&(GHz)	 	 &(mK)&  (\kms)& (mK \kms) &cm$^{-2}$&cm$^{-2}$\\\hline
HNC		&  90.664	&  4.20	 &  0.65	&  20.32&2.0$\times$\pow{11}&1.6$\times$\pow{11}\\
DCN\:$*$	&  144.827	&  9.50	 &  0.40	&  36.05&7.0$\times$\pow{10}&6.0$\times$\pow{10}\\
		&  217.238	&  8.00	 &  0.27	&  24.94&2.3$\times$\pow{10}&2.5$\times$\pow{10}\\
CCH		&  87.316	&  6.20	 &  0.67	&  30.45&1.9$\times$\pow{13}&1.4$\times$\pow{13}\\
DCO$^+$		&  144.077	&  9.20	 &  0.40	&  34.91&6.0$\times$\pow{10}&4.6$\times$\pow{10}\\
N$_2$H$^+$	&  93.171	&  6.50	 &  0.63	&  30.96&8.0$\times$\pow{12}&6.0$\times$\pow{12}\\
HC$_3$N		&  90.979	&  4.20	 &  0.64	&  20.16&3.9$\times$\pow{11}&2.9$\times$\pow{11}\\
		&  136.464	&  7.20	 &  0.43	&  28.33&1.7$\times$\pow{11}&2.4$\times$\pow{11}\\
c-C$_3$H$_2$	&  217.822	&  8.00	 &  0.27	&  24.94&1.0$\times$\pow{11}&1.2$\times$\pow{11}\\
		&  251.311	&  7.20	 &  0.23	&  20.72&8.8$\times$\pow{10}&1.2$\times$\pow{11}\\
p-H$_2$S	&  216.710	&  8.60	 &  0.27	&  26.81&5.4$\times$\pow{12}&2.7$\times$\pow{13}\\
o-H$_2$S	&  204.140	&  15.00 &  0.28	&  47.62&1.6$\times$\pow{16}&5.2$\times$\pow{16}\\
HCS$^+$		&  85.348	&  5.60	 &  0.69	&  27.91&1.8$\times$\pow{12}&1.2$\times$\pow{12}\\
		&  256.027	&  10.00 &  0.23	&  28.77&9.7$\times$\pow{10}&1.1$\times$\pow{11}\\
HCO\:$\diamond$	&  255.341	&  10.00 &  0.23	&  28.77&1.0$\times$\pow{12}&1.4$\times$\pow{12}\\
		&  265.348	&  16.00 &  0.22	&  45.03&3.2$\times$\pow{12}&4.5$\times$\pow{12}\\
SiO		&  86.846	&  6.00	 &  0.67	&  29.47&4.1$\times$\pow{11}&3.2$\times$\pow{10}\\
		&  217.104	&  8.50	 &  0.27	&  26.50&3.5$\times$\pow{10}&4.9$\times$\pow{10}\\
CH$_3$OH	&  252.485	&  11.00 &  0.23	&  31.65&2.4$\times$\pow{13}&1.4$\times$\pow{14}\\
	\hline                                   
	\end{tabular}
\newline
\noindent
{\footnotesize
{\bf Notes}. $^1$ The detection limit was calculated assuming a line width of $\Delta v$ $\sim$4 \kms. The upper limits on the column densities was calculated with MADEX for a source size of $\sim$14 arcsec (2000 AU at 145 pc), a H$_2$ density of 3.5$\times$\pow{6}cm$^{-3}$ for two temperatures: 50 and 30K. $*$ We used the collisional rates of HCN, $\diamond$ Assuming LTE.}
\end{table}

\subsection{Comparison between IRAM 30-m and Plateau de Bure}
\label{Comp_IRAM-30m-PdBI}
\begin{figure}[b!]
 	\centering
 	\includegraphics[scale=0.72]{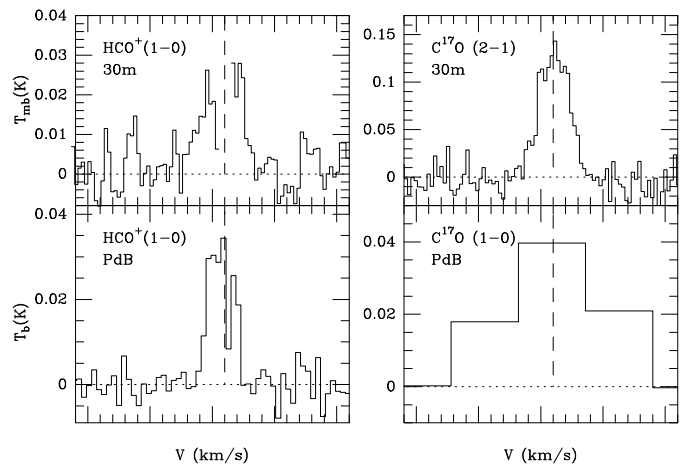}
  	\caption{Comparison between single-dish IRAM 30-m spectra (upper panels)
and Plateau de Bure 		interferometric spectra (lower panels).
	To compare properly both sets of spectra, PdB spectra have been
convolved to the 		single-dish beam size of the corresponding
transition. For the HCO$^+$ 1$\rightarrow$0, the HPBW of the convolving gaussian
is 27.0$\arcsec$ x 27.2$\arcsec$ and for the C$^{17}$O 1$\rightarrow$0,
9.4$\arcsec$ x 9.2$\arcsec$ (to obtain the same angular resolution of the 30-m
C$^{17}$O 2$\rightarrow$1 line). The conversion factor between Jy and K is 7K =
1 mJy/beam for HCO$^+$ 1$\rightarrow$0 and 5K = 1 mJy/beam for C$^{17}$O
1$\rightarrow$0.}
        \label{Fig:comp30m-PdB}
\end{figure}

In Fig. \ref{Fig:comp30m-PdB} we show the HCO$^+$ 1$\rightarrow$0 and C$^{17}$O
1$\rightarrow$0 lines.
The availability of PdB maps allows a comparison between interferometric and
single-dish data. 
To make this comparison, the interferometric maps have been convolved
to the single-dish beam size of the corresponding transition. For the C$^{17}$O
1$\rightarrow$0 line, the temperature measured with the PdB is about four times
less intense than obtained with the IRAM-30 m telescope in the
2$\rightarrow$1 line, which is consistent with optically thin emission. For the 
HCO$^+$ 1$\rightarrow$0 line, all the flux measured with the
IRAM 30-m is recovered by the PdBI, proving that the detected emission comes
from the circumstellar disk.

Looking at the HCO$^+$ 1$\rightarrow$0 spectra of Fig. \ref{Fig:comp30m-PdB}, we
note that the line profile is doubled peaked, which is characteristic of emission from a
rotating disk. However, the line profile seen by the interferometer seems to
peak at a slightly blue-shifted velocity with respect to the IRAM 30-m spectrum.
We believe that this is due to the poor S/N in the PdBI spectrum, which means that the peak velocity and the other line parameters are affected by large
uncertainties.

%%%%%%%%%%%%%%%%%%%%%%%%%%%%%%%%%%%%%%%%%%%%%%%%%%%%%%%%%%%
%		               )\._.,--....,'``.      	%%%
%%	 .b--.    	      /;   _.. \   _\  (`._ ,.   %%
%%%	`=,-,-'~~~   	     `--- (,_..'--(,_..'`-.;.'    % 
%%%%%%%%%%%%%%%%%%%%%%%%%%%%%%%%%%%%%%%%%%%%%%%%%%%%%%%%%%%
\section{Comparison with other sources}
\label{other_sources}

%The sample
For a complete picture of the chemistry in the AB Aur transition
disk, we compared it with the sample of 12 disks observed by \citet{Oberg2010,
Oberg2011}. Such disks were entirely observed with the SMA telescope, ensuring
the uniformity of observational conditions in the sample. Moreover, the lines
detected toward AB Aur are among the lines detected by \citet{Oberg2010,
Oberg2011} at 1mm in the other sources, which facilitated the comparison. 
The sample covers a wide range of stellar luminosities among the TT and Herbig
Ae stars. It consists of six disks in Taurus and six in the southern sky.
Among these disks, we have several (pre-) transition disks with internal gaps: DM
Tau, LkCa 15, GM Aur, V4046 Sgr, HD 142527, and SAO 206462. 
%La tabla
Table \ref{table:others_disks} lists the disk types, pT or T for pre- or
transition disk respectively, and F for full disks. 
The CN 2$\rightarrow$1/HCN 3$\rightarrow$2, HCN
3$\rightarrow$2/HCO$^+$3$\rightarrow$2, and the H$_2$CO
3$_{03}$$\rightarrow$2$_{02}$/HCO$^+$ 3$\rightarrow$2  integrated intensity line
ratios are listed in the other columns. 
%La comparacion
Carrying out a comparison between several sources is not direct, since there are many
uncertainties because the position angles and inclinations affect which parts
of the disk are observed, hence our picture of the chemistry. Disk masses are
poorly constrained, since they are derived from dust emission, assuming the
canonical interstellar dust/gas ratio value of 1/100. The actual dust/gas ratio
may be variable among the sources and also be affected by coagulation and
photoevaporation. The difference in disk masses may affect the relative
abundances of different species.
	         
\begin{table*}[]
	\caption{Disk list of the sample of \citealp{Oberg2010, Oberg2011}, and
AB Aur with their corresponding integrated line intensity ratios of the
transitions;
CN 2$\rightarrow$1/HCN 3$\rightarrow$2, HCN 3$\rightarrow$2/HCO$^+$
3$\rightarrow$2, and H$_2$CO 3$_{03}$$\rightarrow$2$_{02}$/HCO$^+$
3$\rightarrow$2.}  
	\label{table:others_disks}      
	\centering                         
	\begin{tabular}{llcccccccc}       
	\hline\hline     
	     &\multirow{2}{0.5cm}{\small Disk \\type}&L$_*$&
\multirow{2}{1cm}{$\frac{\scriptsize \rm CN\:2\rightarrow1}{\scriptsize \rm HCN 
\:3\rightarrow2}$} 
&  &
\multirow{2}{1cm}{$\frac{\scriptsize \rm HCN\:3\rightarrow2}{\scriptsize \rm 
HCO^+\:3\rightarrow2}$}&	       &
\multirow{2}{1.3cm}{$\frac{\scriptsize \rm H_2CO
3_{03}\rightarrow2_{02}}{\scriptsize \rm  HCO^+\:3\rightarrow2}$} &	&\\
             Sources	& &\Lsun&&error &&error  && error&Refs.\\\hline
DM Tau	     &pT  &0.3	  &1.62   &0.09  &0.55   &	0.03   &0.07	& 0.01 &a\\
AA Tau	     &F	  &0.59	  &2.98   &0.81  &0.25   &	0.08   &0.07	& 0.04 &b\\
LkCa 15      &pT  &1.2	  &1.23   &0.07  &1.06   &	0.08   &0.13	& 0.02 &a\\
GM Aur	     &pT  &0.93	  &0.76   &0.13  &0.25   &	0.03   &0.11	& 0.02 &a\\
CQ Tau	     &F	  &8	  &>0.44  &      &<0.67  &	       &<0.28	&      &c\\
MWC 480      &F	  &11.5	  &1.41   &0.17  &0.50   &	0.07   &<0.06	&      &d\\
IM Lup	     &F	  &1.3	  &1.01   &0.11  &0.36   &	0.04   &0.05 	& 0.01 &e\\
AS 205	     &F	  &4	  & -     &      &<0.14  &	       &<0.06	&      &f,g\\
AS 209	     &F	  &1.5	  &1.65   &0.18  &0.50   &	0.05   &<0.05	&      &g,h\\
V4046 Sgr    &pT  &0.5	  &1.22   &0.05  &0.87   &	0.03   &0.09	& 0.02 &i\\
SAO 206462   &T	  &8	  &1.00   &0.50  &0.49   &	0.22   &<0.25	&      &j\\
HD 142527    &T	  &69	  & -     &      &<0.10  &	       &0.03	& 0.02 &j\\
AB Aur       &T   &38	  &0.44   &0.10  &0.24   &	0.02   &0.12	& 0.01 &k\\
	\hline                                   
	\end{tabular}
\newline
\noindent
{\footnotesize
{\bf Notes}. (a) \citet{Espaillat2010}; (b) \citet{White2001}; (c)
\citet{Mannings1997}; (d) \citet{Simon2000};
(e) \citet{Hughes1994}; (f) \citet{Prato2003}; (g) \citet{Andrews2009}; (h)
\citet{Herbig1988};
(i) \citet{Quast2000}; (j) \citet{Garcia2006}; (k) \citet{DeWarf2003}}
\end{table*}

\begin{figure*}[t!]
	\centering
	\includegraphics[scale=0.9]{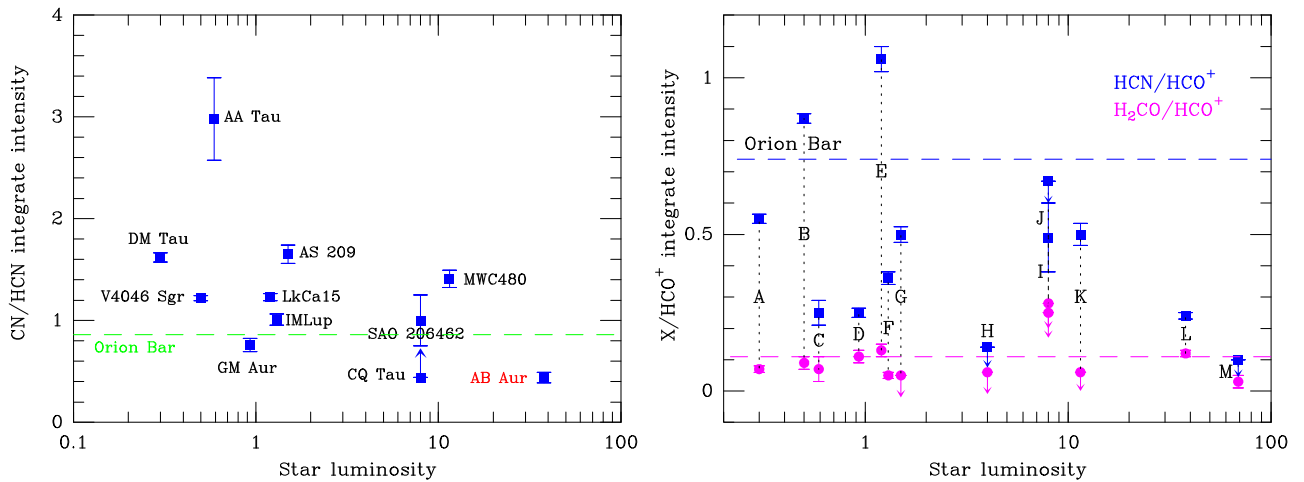}
	 \caption{Integrated line intensity ratios and upper and lower limits
for CN 2$\rightarrow$1/HCN 3$
\rightarrow$2 ({\rm \emph{left}}) and HCN 3$\rightarrow$2/HCO$^+$
3$\rightarrow$2 and H$_2$CO 3$_{03}$$\rightarrow$2$_{02}$/HCO$^+$
3$\rightarrow$2 ({\rm \emph {right}}) transitions, as a function of the stellar
luminosity. \emph {Horizontal dashed lines} indicated the corresponding values
for the Orion Bar (J. R. Goicoechea, private communication). The letters
correspond to each of the sources. (A) DM Tau; (B) V4046 Sgr; (C) AA Tau; (D) GM
Aur; (E) LkCa 15; (F) IM Lup; (G) AS 209; (H) AS 205; (I) SAO 206462; (J) CQ Tau; (K) MWC
480; (L) AB Aur; (M) HD 142527.}
	\label{Fig:ratio}%
\end{figure*}

%CN/HCN
Among the molecular species detected in AB Aur, we have HCN and CN. They are
particularly interesting because they trace different physical conditions in the
photodissociation regions (PDRs) and because their abundances are closely
related HCN may be easily photodissociated by UV radiation from the star to
yield CN (\citealp{Fuente1993, Bachiller1997, Boger2005}).

Both molecules, CN and HCN, share the same chemistry and are expected to come 
from the same region, although the CN/HCN ratio can vary (see Fig. \ref{Fig:map_mole}). 
Furthermore, the CN 2-1 and HCN 3-2 lines have similar critical densities and therefore 
trace regions with similar physical conditions. Assuming optically thin emission, their 
integrated line intensities are proportional to the number of molecules of each species. 
With all these assumptions, the CN 2-1/HCN 3-2 line ratio is a good tracer of the CN/HCN 
abundance ratio.

In Fig. \ref{Fig:ratio} we show the integrated intensity line ratios and upper
and lower limits for CN 2$\rightarrow$1/HCN 3$\rightarrow$2, HCN
3$\rightarrow$2/HCO$^+$3$\rightarrow$2, and H$_2$CO
3$_{03}$$\rightarrow$2$_{02}$/HCO$^+$ 3$\rightarrow$2 ratios. 
The CN/HCN ratios vary by factor 4 between the DM Tau and AB Aur. AA Tau is the only significant 
outlier in terms of CN/HCN flux ratio, which has
a value of 2.9. The averaged value of the CN/HCN ratio in the sample is $\sim$1.1. This value
is similar to the one measured in the Orion Bar PDR $\sim$0.85 (J. R. Goicoechea, private communication) 
(see Fig. \ref{Fig:ratio}). 
CN and HCN are more abundant in the illuminated disk surface (see Fig. \ref{Fig:map_mole}). Chemical 
models show that the CN/HCN ratio is a good tracer of the incident UV field \citep{Fuente1993, Boger2005}, 
taking values of about 1 in dark clouds and increasing to > 3 in PDRs.
The similarity between the CN 2-1/HCN 3-2 line ratio in AB Aur and Orion is consistent with the 
interpretation of CN and HCN emission coming from the PDR formed in the inner part of the disk and the 
illuminated disk surface.

We found a weak trend between CN 2$\rightarrow$1/HCN 3$\rightarrow$2 ratio and
the morphology of the disk. Transition disks with high luminosities have the
lowest values for CN/HCN ratios. 
AB Aur is located with the other transition disks, SAO 206462 and
HD 142527, in the bottom right hand part of the plot (see Fig \ref{Fig:ratio}).

The average CN/HCN abundance ratio is expected to depend on the total area of the illuminated disk 
surface and the intensity of the UV field.
Assuming the same morphology for all the disks, one would expect that the CN/HCN ratio increases 
with the stellar luminosity. However, we find the contrary trend. We interpret that this trend might 
be due to the different disk morphologies. In the transition disks with large inner gaps, the emission 
is dominated by the outer ring, i.e., the region farthest from the central star, where the FUV flux is 
weaker. This effect would be increased if the disks were flatter, because the illuminated surface would be 
drastically reduced.

In Fig. \ref{Fig:ratio} we also plot the HCN
3$\rightarrow$2/HCO$^+$3$\rightarrow$2 and the H$_2$CO
3$_{03}$$\rightarrow$2$_{02}$/HCO$^+$ 3$\rightarrow$2 ratio. The averaged value
of HCN/HCO$^+$ ratio in the sample is 0.43 and for H$_2$CO
3$_{03}$$\rightarrow$2$_{02}$/HCO$^+$ 3$\rightarrow$2 is 0.1. 
We did not find any trend for each of these ratios with the stellar luminosity
or the disk morphology. However, we found a relation between them both. Clearly,
the HCN 3$\rightarrow$2/HCO$^+$3$\rightarrow$2 ratio is higher than H$_2$CO
3$_{03}$$\rightarrow$2$_{02}$/HCO$^+$ 3$ \rightarrow$2 ratio in all the cases.
This is also true for the Orion Bar: their corresponding values are 0.74 for
the HCN/HCO$^+$ and 0.11 for H$_2$CO/HCO$^+$ (J.R. Goicoechea, private
communication).

%%%%%%%%%%%%%%%%%%%%%%%%%%%%%%%%%%%%%%%%%%%%%%%%%%%%%%%%%%%
%		               )\._.,--....,'``.      	%%%
%%	 .b--.    	      /;   _.. \   _\  (`._ ,.   %%
%%%	`=,-,-'~~~   	     `--- (,_..'--(,_..'`-.;.'    % 
%%%%%%%%%%%%%%%%%%%%%%%%%%%%%%%%%%%%%%%%%%%%%%%%%%%%%%%%%%%
\section{Modelization}
\label{Modelization}

\subsection{Physical model}

\begin{table}[b]
\caption{Input parameters used for modeling the AB Aur disk. We varied
the disk mass, the inner radius, the outer radius of the disk, and the radial
index, $\alpha$, of the surface density profile, $\Sigma(R)\propto R^{-\alpha}$.
The disk mass is given in units of solar masses (\Msun) and radii
in astronomical units (AU).}           
	\label{table:models}      
	\centering                         
	\begin{tabular}{c c c c c c}       
	\hline\hline     
		&A	&B	&C	&D	&E	\\\hline
M$_{{\rm disk}}$&0.02	&0.01 	&0.01 	&0.01	&0.01	\\
R$_{{\rm in}}$	&70 	&70	&110	&70	&110	\\
R$_{{\rm out}}$	&1100 	&550	&550 	&1100	&550	\\
$\alpha$	&2.15  	&2.15	&2.15 	&2.15	&1.5	\\
	\hline                                   
	\end{tabular}
\end{table}

For deeper insight into the chemistry of this Ae disk, we carried
out several modeling tasks.
Our model assumes a flared disk with the density profile in the vertical axis
determined by the assumption of hydrostatic equilibrium. 
The density and temperature distributions are shown in Fig. \ref{Fig:map_temp}
The temperature and visual extinction are computed using the
RADMC code \citep{Dullemond2004}, which solves the continuum radiative transfer throughout the
disk.
We use a dust model with the MRN grain size distribution \citep*{Mathis1977}
made of 100\% silicates.
Gas kinetic temperature is higher than dust temperature in the very external
layers on the disk surface, where the photoelectric effect is the main heating
mechanism. However, the millimeter lines studied in this paper come from a
denser and cooler region where gas and dust are expected to be thermally
coupled. In our model we assumed that the gas and dust temperature are equal.

Flared disks are expected around young stars \citep{Meeus2001}.
The input parameters are the disk mass (M$_{\mathrm{disk}}$), the inner and
outer radii (\Rin and \Rout) and the radial index, $\alpha$, of the surface
density profile ($\Sigma(R)\propto R^{-\alpha}$).

As an initial trial, we considered Models A, B, C, and D with the parameters
shown in Table \ref{table:models}. We adopted the disk mass of 0.02 \Msun
derived by \citet{Pietu2005} from continuum 1mm PdB 
observations and allowed it to change by a factor of 2, so within its
uncertainty. The values of \Rin and \Rout are varied between the values found
from $^{12}$CO and those obtained from the rare CO isotopologues and HCO$^+$
(\citealp{Pietu2005, Schreyer2008}, this paper).
The value of $\alpha$ was taken 2.15, in agreement with previous measurements by
\citet{Pietu2005} and \citet{Schreyer2008}.

The parameters adopted in Models A,B,C, and D are based on fitting interferometric observations with a simple disk model.
More recent high spatial resolution observations of CO and $^{13}$CO showed that
the AB Aur disk has a complex structure,  
with a rotating ring that extends to 4$\arcsec$ (560 AU) and a spiral arm
extending up to 7$\arcsec$ ($\approx$1000 AU) (\citealp{Tang2012,Pantin2005}). 
Moreover, according to observations by \citet{Hashimoto2011}, there are two dusty
rings:
one inmate compact ring (R $\approx$ 40 AU) with an inclination angle of
approximately 40$\degree$ relative to the plane of the sky and the outer ring
with R $\approx$ 100 AU and an inclination angle of 27$\degree$. 
\citet{Honda2010} derived a value of $\alpha$ close to 1 for the outer ring. Our
model is too simple to account for this complex structure,
but we added Model E (see Table \ref{table:models}) to account for the lower
radial spectral index measured by \citet{Honda2010}. 

In all the models, the stellar parameters were kept fixed to T$_*$ = 10 000 K,
R$_*$ = 2.5 R$_{\odot}$, M$_*$ = 2.4 M$_{\odot}$ \citep{Schreyer2008}.
and the gas-to-dust mass ratio to 100. 

\subsection{Chemical model}

We used a simple time-dependent chemical model in which the representative
chemical abundances evolve under fixed physical 
conditions (updated version of \citealp{Agundez2008} and \citealp{Fuente2010}).
The chemical model includes the elements H, C, N, O, and S. As initial
abundances we adopt the so-called “low metal” values \citep{Wiebe2003}, with all 
carbon in the form of CO, the remaining oxygen
in the form of water ice, nitrogen as N$_2$, sulfur as CS, and a ionization
fraction of 1$\times$10$^{-8}$(typical of dark clouds, see, e.g., \citealp{Caselli1998, Agundez2013}). 
The gas phase chemical
network is an extension of the one used by \citet{Agundez2008}. 
We also consider adsorption onto dust grains and desorption
processes, such as thermal evaporation, photodesorption, and desorption
induced by cosmic rays (see, e.g., \citealp{Hasegawa1993}). 
The photodissociation rates of H$_2$ and CO are computed including
self-shielding effects. For the stellar FUV field we adopt the
approach described in \citet{Agundez2008}, while in the
case of the interstellar FUV field, we use the shielding coefficients
calculated by \citet{Lee1996}.
The adopted H$_2$ formation rate on the grain surface is 3$\times$10$^{-17}$ cm$^3$
s$^{-1}$.

\begin{figure*}[t]
   \centering
   \includegraphics[scale=1.3]{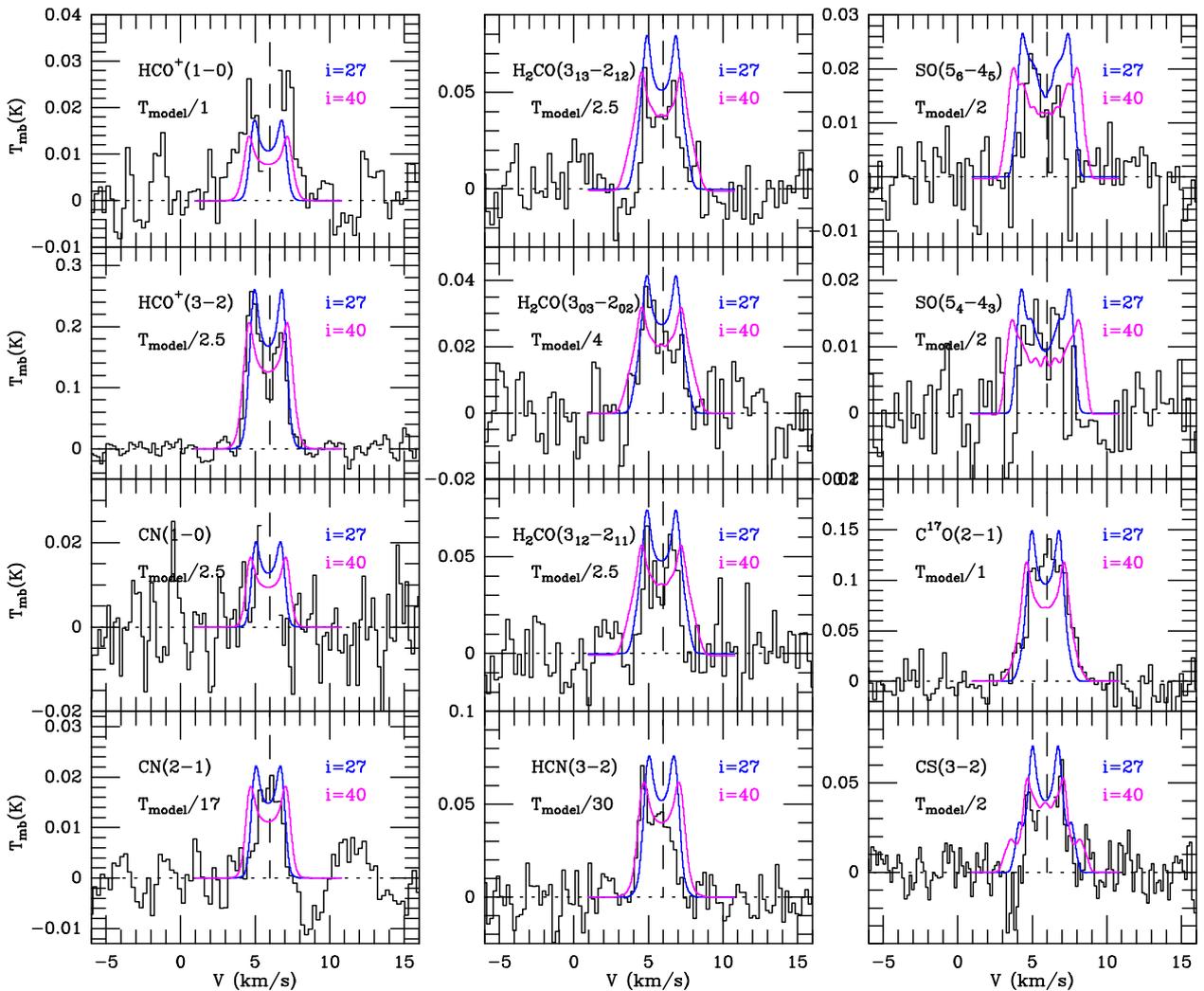}
   \caption{Comparison between modeled spectra and those detected by 30m
telescope toward AB Aur disk. In colors, we have the lines profiles that
correspond to the synthetic line fluxes performed by MCFOST code. The
\emph{blue} and \emph{magenta} lines correspond to the same model with
inclination angles for the disk of 27$\degree$ and 40$\degree$, respectively.
The vertical dash line indicates the ambient cloud velocity, V$_{\rm{lsr}}$ = 5.9
\kms~ \citep{Duvert1986}. All the observational spectra show the double-peak
shape that is characteristic of a rotating disk. All modeled lines agree with the observations within a factor of 2. HCN 3$\rightarrow$2 and CN
2$\rightarrow$1 present the largest differences between the predicted
intensities and the observations, 30 and 17 times larger than observations,
respectively.}
\label{figura:detecciones_modelE}
\end{figure*}

Using the chemical composition computed by our model, we performed the 3D radiative transfer code
MCFOST \citep{Pinte2006, Pinte2009} to compute
the synthetic molecular emission maps. MCFOST uses the chemical abundances and
kinetic temperature to compute the level populations using a Monte Carlo
method. 
We followed an algorithm similar to the one published by \citet{Hogerheijde2010},
separating the contribution of the local and external radiation fields in each
cell.
The synthetic emission maps and spectra are then produced by a ray-tracing
algorithm that formally integrate the source function computed by the Monte Carlo
method.
We iterated this procedure several times over a range of position angles from 
0$^{\circ}$ to 40$^{\circ}$, relative to the plane of the sky. Finally, we
convolved the resulting map with the beam of Pico de Veleta to compare with
observations. The results are shown in Fig. \ref{figura:detecciones_modelE} and
in Figs. \ref{Fig:modelAB} and \ref{Fig:modelCD}.

\subsection{Comparison between model and observations}
\label{Comparison}

Our models from A to E are reasonable approximations of the AB Aur disk (see
Figs. \ref{figura:detecciones_modelE}, \ref{Fig:modelAB} and \ref{Fig:modelCD}).
Model A corresponds to a disk with mass of 0.02 \Msun and a large outer radius
of 1100 AU. This model overestimates the observed
intensities and the line widths for all transitions. In Model B we used the same parameters as in
Model A, but we changed the outer radius to 550 AU and the disk mass to 0.01
\Msun. By decreasing and redistributing the mass in a smaller radius, we observed
a decrease in the line intensities, but there is no change in the width of the
profiles. In Model C we increased the size of internal radius, the synthesized
spectra fit the width of the lines best, and the line profiles are more similar
to those of the observed spectra. Still, under the assumption of a mass of 0.01
\Msun and when testing the same other parameters as in Model A, we noted the effect
of decreasing the mass in the disk in Model D. The lines are still wider than
observed, although the intensities are adjusted relatively well within a factor
of 2 for most of the transitions. 

\begin{figure}[b!]
	\centering
	\includegraphics[scale=0.6]{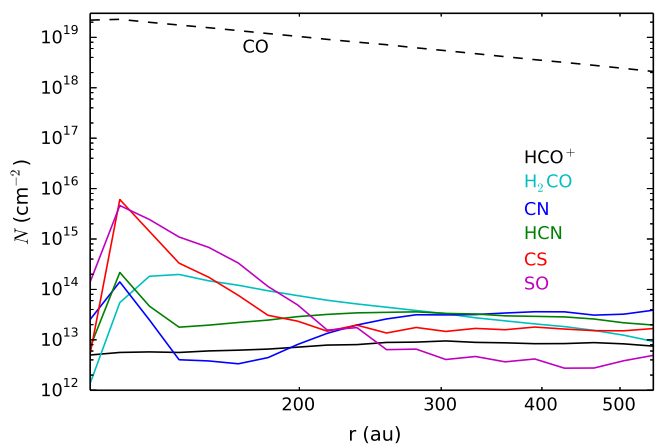}
	\caption{Vertical column densities of several molecules as a function of
radius at 2.5 Myr as 		calculated for the chemical model E.}
        \label{Fig:densi_colum}
\end{figure}

Finally, Model E is like the previous Model C, but it changes the exponent of the
radial surface density law from 2.15 to 1.5. The best results were obtained for
this model, which corresponds to the values M$_{\rm{disk}}$=0.01\Msun, \Rin=110
AU, \Rout=550 AU, and a surface density radial index $\alpha$=1.5 with an
inclination angle of i=27\degree.
Almost all modeled lines agree with the observations within a factor
of 2, and the modeled profiles successfully reproduce the double peak profiles
from the disk emission. The exceptions are the HCN 3$\rightarrow$2 that is
predicted with an intensity around 30 times higher than the observed ones and
the CN 2$\rightarrow$1 with a predicted intensity that is 17 times higher.
This large difference was highlighted in the eye inspection during the data
analysis. We noted that HCN 3$\rightarrow$2 and CN 2$\rightarrow$1 were
especially weak.
This peculiarity in the emission is also observed as a common trend in
transition disks.

We call that our chemical model does not consider surface reactions that could
affect the HCN and CN chemistry. However, our model gives vertical HCN column
densities of $\sim$~\pow{14} cm$^{-2}$, in agreement with other more complex
models that include surface chemistry (see, e.g., \citealp{Walsh2010}).
Therefore, we think that the discrepancies between observations and our
predictions are more likely due to the assumed morphology of the AB Aur disk. 

In Fig. \ref{Fig:densi_colum} we show the HCN and CN vertical column densities as a function of the
disk radius. Both species have peaks close to the inner radius of 110 AU.
Subsequently, they decrease to a minimum value around 200 AU and increase again
for larger radii because of the PDR formed on the surface of the flared disk. In
the case of a flat disk, the HCN and CN column densities would be lower for
large radii and in better agreement with our observations.
This could also be the cause of the low HCN and CN intensities in other
transition disks.

In our model the SO synthesized lines agree with the observations by
a factor of 2. SO emission has been detected associated to disks of young low-mass Class I protostars \citep{Yen2014}.
The youth of the AB Aur disk could be key for a higher SO abundance. 
Moreover, the AB Aur disk is warmer than those associated with low-mass stars and, 
unlike to these disks, which present a standard abundance of CO without signs 
of heavy depletion (\citealp{Pietu2005,Schreyer2008}).

\begin{table}[]
	\caption{Initial abundances in Models M0, M1, and M2$^1$.}           
	\label{table:abundances}      
	\centering                         
	\begin{tabular}{p{1cm} c c c}       
	\hline\hline     
	 & M0          &       M1     &          M2   \\                     
Species  &abun/H$_{\textrm {tot}}$&abun/H$_{\textrm {tot}}$ &abun/H$_{\textrm {tot}}$\\\hline 
\multicolumn{4}{c}{gas phase}\\\hline
 H       &  0.000E+00  &   2.1414E-04 &   1.6066E-04  \\
 H$_2$   &  0.500E+00  &   4.9989E-01 &   4.9992E-01  \\
 He      &  0.085E+00  &   8.5000E-02 &   8.5000E-02  \\
 C       &  0.000E+00  &   1.2132E-05 &   4.5720E-08  \\
 O       &  0.000E+00  &   1.1569E-04 &   8.6790E-06  \\
 N       &  0.000E+00  &   2.1795E-05 &   5.1383E-07  \\
 S       &  0.000E+00  &   8.6586E-08 &   1.5544E-08  \\
 OH      &  0.000E+00  &   5.3736E-08 &   4.3489E-07  \\
 H$_2$O  &  0.000E+00  &   2.7982E-07 &   1.3307E-07  \\
 O$_2$   &  0.000E+00  &   1.1912E-07 &   6.3504E-06  \\
 CO      &  7.860E-05  &   5.9802E-05 &   4.3019E-05  \\
 CO$_2$  &  0.000E+00  &   1.0848E-08 &   6.9679E-08  \\
 H$_2$CO &  0.000E+00  &   1.0260E-08 &   8.1825E-10  \\
 NH$_3$  &  0.000E+00  &   2.6452E-09 &   4.6800E-08  \\
 N$_2$   &  1.235E-05  &   1.3839E-06 &   8.6479E-06  \\
 CN      &  0.000E+00  &   1.4022E-08 &   4.8573E-09  \\
 HCN     &  0.000E+00  &   1.2226E-08 &   6.2810E-09  \\
 HNC     &  0.000E+00  &   6.9881E-09 &   5.4021E-09  \\
 CS      &  9.140E-08  &   2.3049E-09 &   2.3040E-10  \\
 SO      &  0.000E+00  &   3.3169E-10 &   1.0105E-08  \\\hline
 \multicolumn{4}{c}{ice}\\\hline
 CH$_4$ &  0.000E+00  &   2.3428E-08 &   6.1490E-08  \\
 H$_2$O &  1.014E-04  &   6.7127E-08 &   1.0762E-06  \\
 O$_2$  &  0.000E+00  &   1.0774E-08 &   3.8826E-05  \\
 CO     &  0.000E+00  &   3.7799E-06 &   3.5103E-05  \\
 CO$_2$ &  0.000E+00  &   2.2142E-09 &   2.1759E-07  \\
 H$_2$CO&  0.000E+00  &   2.6391E-09 &   5.4471E-09  \\
 NH$_3$ &  0.000E+00  &   4.8238E-10 &   1.4827E-07  \\
 N$_2$  &  0.000E+00  &   3.7448E-08 &   3.0605E-06  \\\hline                                   
\end{tabular}

\noindent
\hspace{-1.8cm}$^1$Notation: 2.1414E-04= 2.1414$\times$10$^{-04}$
\end{table}

In general, there is a discrepancy of almost a constant factor of 2 between our
model predictions and observed line intensities. This factor may be due to a
different gas/dust ratio from the standard value, and it could even vary within the
disk. This could produce a decrease in the intensity of the lines coming from
the gas-deficient region.
It is definitely not the unique explanation, but taking the 
large uncertainties of our simple model into account, we think it is not worthwhile discussing this disagreement further.
We call that the gas/dust ratio is one of the least known parameters in all the disk models.

The shortcoming of time-dependent models is that the predictions depend on the 
initial conditions. In the following section, we discuss the impact of such initial conditions on our results.

\subsection{Dependence on initial conditions}

\begin{figure*}[]
	\centering
	\includegraphics[scale=1.3]{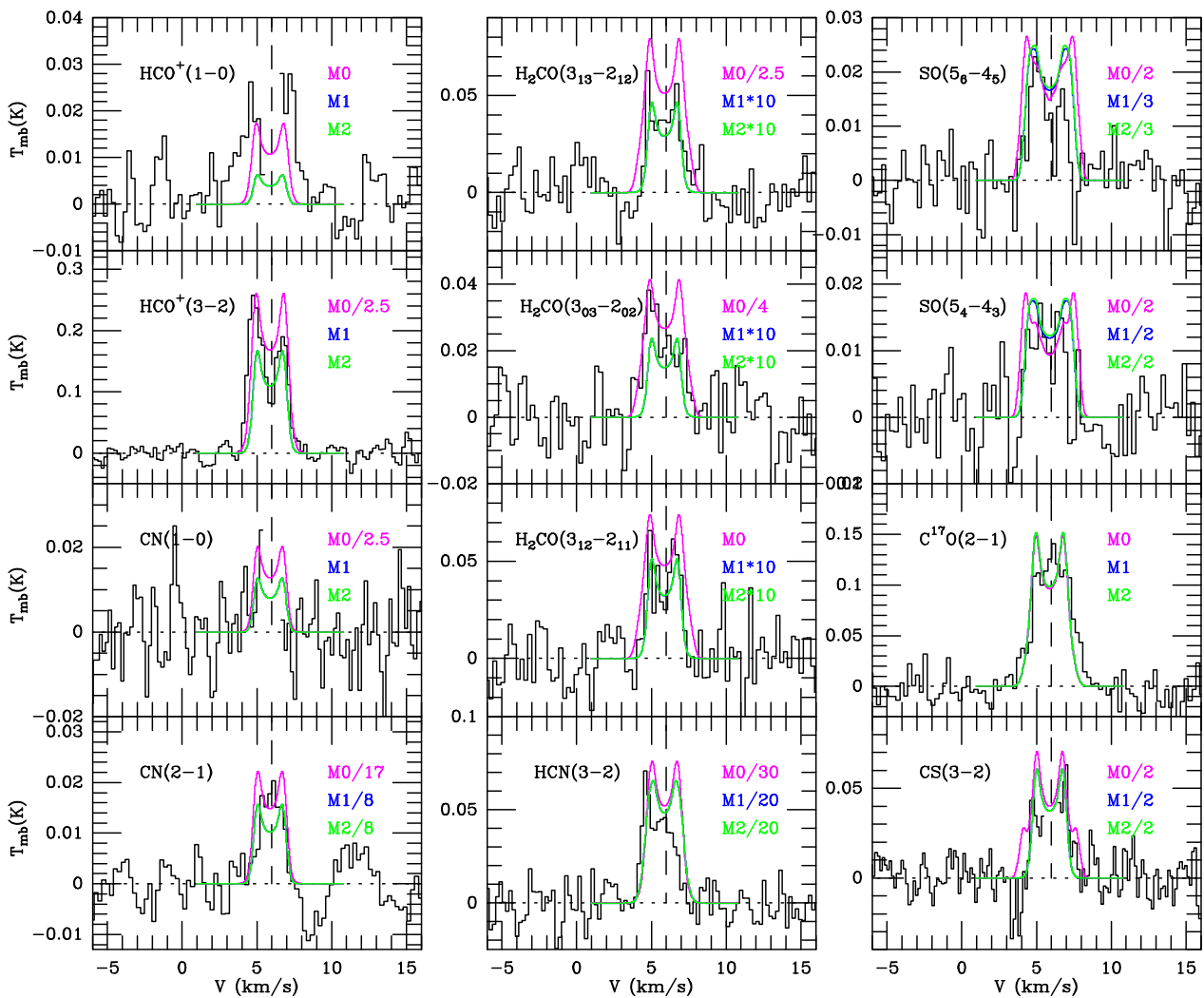}
	\caption{Detected spectra toward AB Aur overlaid with the
synthesized transition lines obtained with the radiative transfer code MCFOST
for models M0, M1, and M2 for an angle of 27$\degree$ for the physical parameters of model E.}
        \label{Fig:MO-M1-M2}
\end{figure*}

To analyze the impact of the initial condition on our model, we used a trial of
initial abundances: M0 (the actual model used up to now in this work), M1 and M2. 
The values for each set of abundances are shown
in Table \ref{table:abundances}. Model M0 corresponds to a standard solar elemental 
composition corrected with a depletion around 100 for Na, Si, Mg, and Fe 
(see \citealp{Wiebe2003, Semenov2004}). In this model, all the N is in the form of N$_2$, 
the C in CO, the O in water ice, and all the S in CS. (We chose this option as the limit 
case to verify the model's ability to form SO.)
M1 and M2 initial abundances values correspond to the values of a dark cloud 
(n$_\textrm{H}$=2$\times$ \pow4cm$^{-3}$, T(K)=10K) at a time of 10$^5$ and 10$^6$ years, 
respectively.

Figure \ref{Fig:MO-M1-M2} shows the comparison between the observed spectra and the
profiles synthesized with the radiative transfer code MCFOST
for models M0, M1, and M2 and for an angle of 27\degree. Models M1 and M2 show practically the same line
profiles for all transitions.
As an outstanding result, the CS and SO predicted line profiles do not vary with
the initial conditions, supporting the robustness of our results. 

In the case of CN and HCN, the predicted intensities are lower 
than in M0, but remain more intense than the observations.
Thus, varying the initial conditions does not improve our results.
This discrepancy between models and observations may be due to a different physical 
structure (see Sect. \ref{Comparison}) or to the uncertainties in the nitrogen chemistry. 
For instance, the values of elemental abundances, the ortho-to-para H$_2$ ratio, 
the binding energies, and time-dependent effects could affect our results (see, e.g., 
\citealp{LeGal2014, Schwarz2014}). However, a detailed study of these effects is beyond the scope of this work.

Models M1 and M2 fail to predict the abundances of formaldehyde by a factor of 10, because the formaldehyde emission mainly comes from the midplane where the
chemistry processes are slow and the molecular abundances are far from the equilibrium. 
In Models M1 and M2, there is, first, a period of very early
formation of molecules, followed by a time of destruction of the chemical
complexity (between a few 10$^5$ and 10$^7$ yr), and then the formaldehyde abundance 
increases again to reach the equilibrium value at 10$^8$ yr. The steady-state
abundance is more like the result of our M0 model. 

Among the 
studied molecules, only the abundance of H$_2$CO strongly depends on the
assumed initial conditions.

%%%%%%%%%%%%%%%%%%%%%%%%%%%%%%%%%%%%%%%%%%%%%%%%%%%%%%%%%%%
%		               )\._.,--....,'``.      	%%%
%%	 .b--.    	      /;   _.. \   _\  (`._ ,.   %%
%%%	`=,-,-'~~~   	     `--- (,_..'--(,_..'`-.;.'    % 
%%%%%%%%%%%%%%%%%%%%%%%%%%%%%%%%%%%%%%%%%%%%%%%%%%%%%%%%%%%
\section{Summary and conclusions}
\label{Summary}

We present in this paper the molecular survey toward the prototypical disk
around AB Aur. Our results and conclusions can be summarized as follow.

   \begin{enumerate}

	\item We detected the 1$\rightarrow$0 and 2$\rightarrow$1 rotational
transitions of $^{12}$CO and its isotopologues: $^{13}$CO, C$^{17}$O, and
C$^{18}$O.
We also detected the lines HCO$^{+}$ 1$\rightarrow$0 and 2$\rightarrow$1,
H$_2$CO 3$_{03}$$\rightarrow$2$_{02}$, 3$_{13}$$\rightarrow$2$_{12}$, and
3$_{12}$$\rightarrow$2$_{11}$, SO 5$_4$$\rightarrow$4$_3$,
5$_6$$\rightarrow$4$_5$, CS 3$\rightarrow$2, HCN 1$\rightarrow$0,
3$\rightarrow$2, and CN 1$\rightarrow$0 and 2$\rightarrow$1.
Based on the line profiles and on the agreement between the velocity ranges, we
interpreted that the emission from all these lines comes from the disk. 
We therefore confirmed the previous detection of SO in AB Aur by Fuente et al.
(2010).
  	\item Some species were not detected in our survey, even if they were
observed in other disks, such as HNC, DCN, C$_2$H, DCO$^+$, N$_2$H$^+$, HC$_3$N, and
C$_3$H$_2$. In general, these species have weaker lines whose emission could be
below our sensitivity limit.

	\item Carrying out a comparison with other 12 disks \citep{Oberg2010,
Oberg2011} that cover a broad luminosity range and disk evolutionary stages, we
found that AB Aur presents the typical chemical signature of a transition disk.
Transition disks have lower CN 2$\rightarrow$1/HCN 3$\rightarrow$2 ratios than
full disks. Moreover, the HCN 3$\rightarrow$2/HCO$^+$ 3$\rightarrow$2 and
H$_2$CO 3$\rightarrow$2/HCO$^+$ 3$\rightarrow$2 ratios are similar.
\end{enumerate}	

We modeled the line profiles using the chemical code by \citet{Agundez2008} 
and the 3D radiative transfer code MCFOST \citep{Pinte2006,Pinte2009}. Our model assumes a flared disk in hydrostatic equilibrium with a
radial surface density distribution described by $\alpha$ ($\Sigma(R)\propto
R^{-\alpha}$). The best results were obtained for a physical structure: 
M$_{\rm{disk}}$=0.01\Msun, \Rin=110 AU, \Rout=550 AU, a surface density coefficient ($\alpha$=1.5),
and the initial abundances M0.
Almost all modeled lines, in particular SO and CS, agree with the observations
within a factor of 2, and the modeled profiles successfully reproduce the double-peak profiles from the disk emission. The only exception is the HCN
3$\rightarrow$2 that is predicted with an intensity around 30 times higher and
the CN 2$\rightarrow$1 with a predicted intensity of 17 times higher than the
observed ones, respectively.

We carried out a set of modeling tasks with different initial abundances to study the 
influence of the initial conditions on our results. We took the limiting cases in which all 
the S is in CS (model M0) and all the S is in SO (model M2). We found that the initial 
conditions do not affect our result, supporting its robustness. The CN and HCN synthetic spectra 
always remain more intense than the observations, thus varying the initial conditions do not significantly 
improve our results. Only the H$_2$CO abundance strongly depends on the initial conditions.

Our model cannot account for the full complexity of this transition disk and
must be considered as a guide to interpreting the observations. For instance, clumpyness in the disk ring would allow UV
photons to penetrate deeper into the disk and photodissociate some molecules.
A flatter disk morphology would lead to different chemical abundances.
Recent observations in centimeter radio emission, carried out by
\citet{Rodriguez2014}, have shown that there is a collimated and ionized outflow in
the inner regions of AB Aur disk. 
Possible shocks associated with this compact ionized outflow could modify the
gas chemistry in the interaction regions.
Recent ALMA observations toward another transition disk HD142527 have traced
HCO$^+$ emission inside the cavity, and it has been interpreted as accretion
flows onto the star \citep{Casassus2013}. 
Finally, dust properties and the gas/dust ratio in transition disks could be
different from the standard values (see, e.g., \citealp{Marel2014}).
Future interferometric observations would help to determine where the
emission of different molecules are coming from and distinguish among all these scenarios.

\vspace{0.5cm}
\begin{acknowledgements}
      We acknowledge the financial support of CONACyT, M\'exico, and Spanish MINECO for
funding support under grants CSD2009-00038,  FIS2012-32096 and AYA2012-32032.
\end{acknowledgements}

\bibliography{AB_Aur_final}
\bibliographystyle{aa}

\newpage
\begin{appendix}
\section{Figures}
\label{app1}

\begin{figure*}
	\centering
	\subfigure{\includegraphics[scale=0.36]{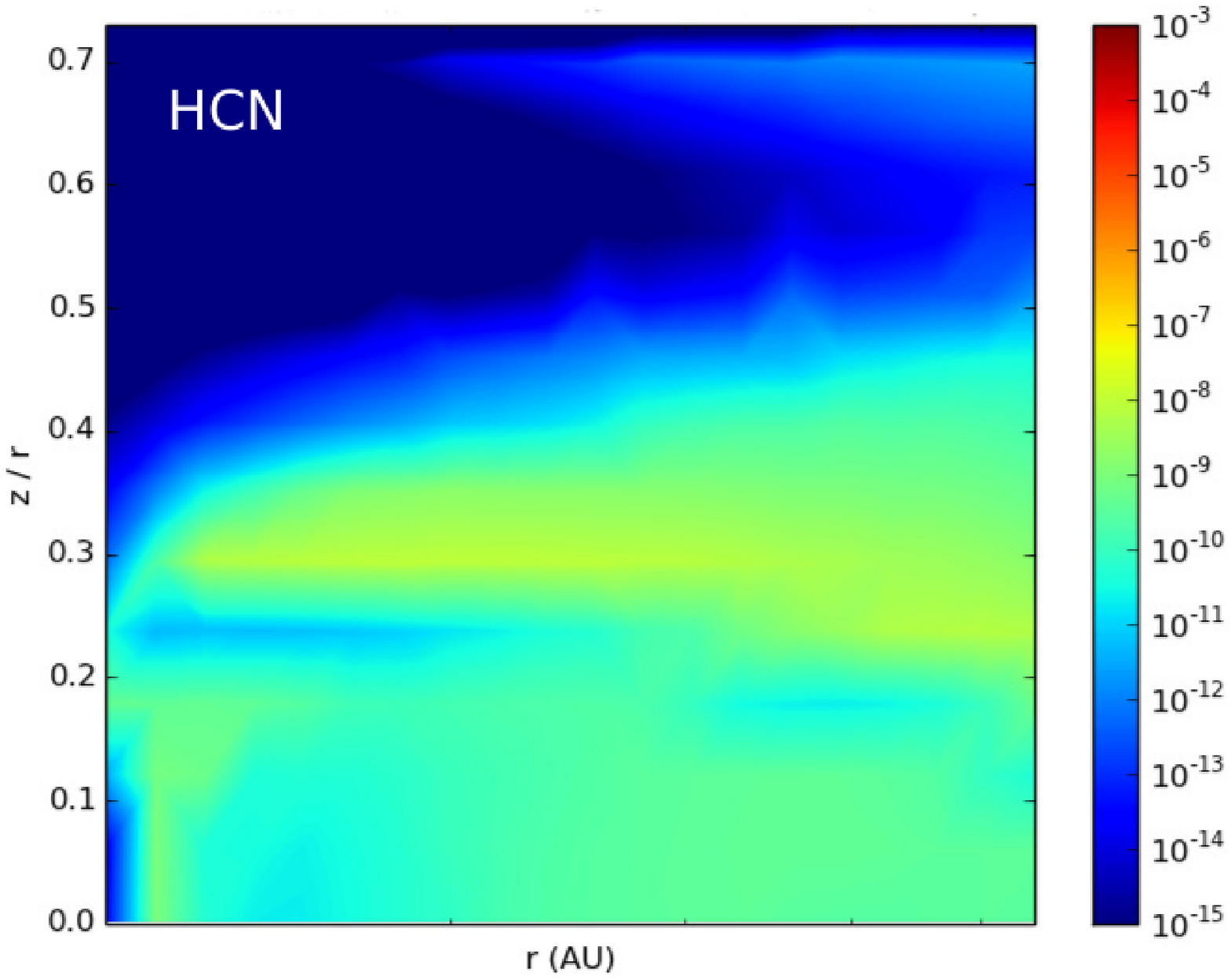}}
	\subfigure{\includegraphics[scale=0.36]{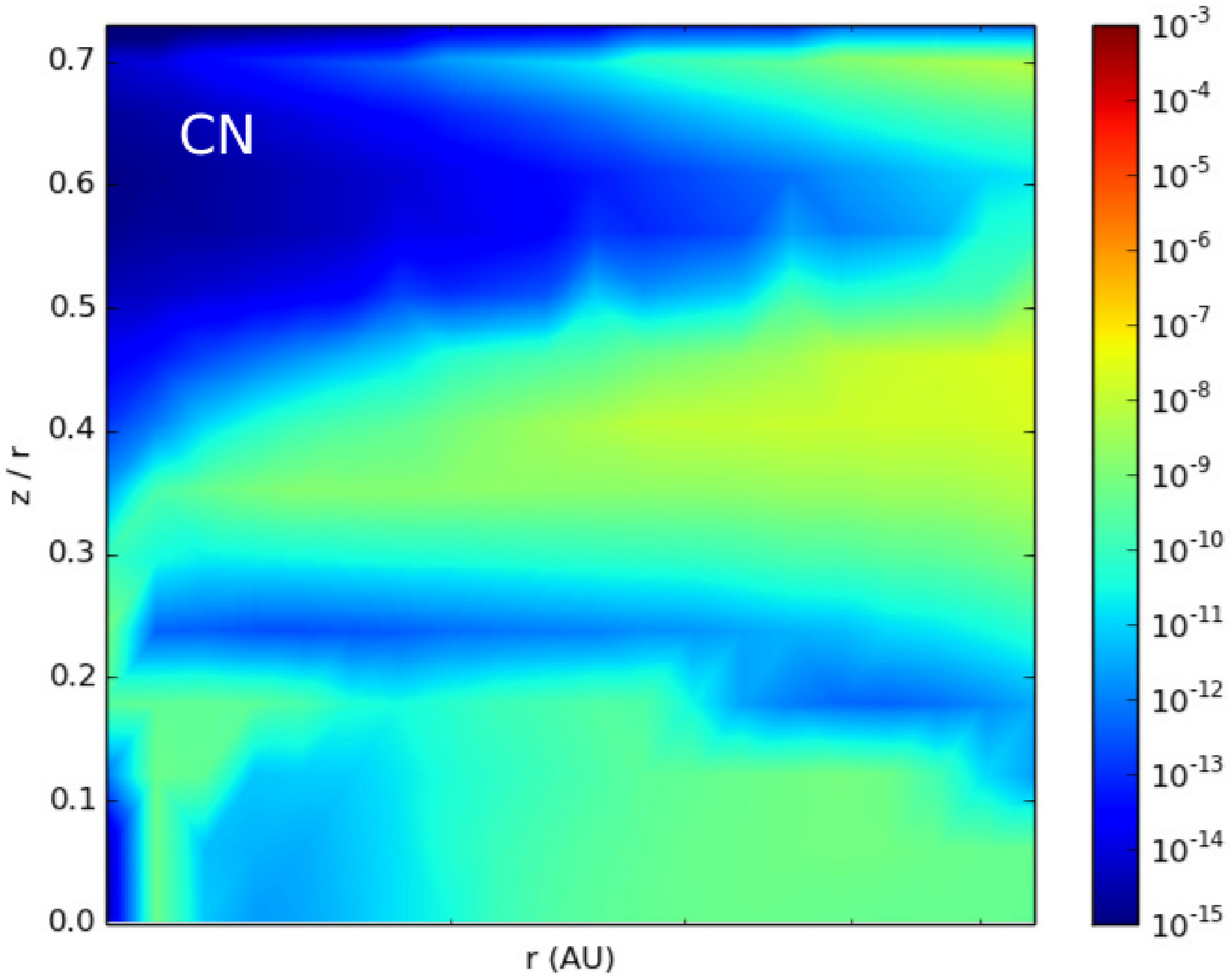}}
	\subfigure{\includegraphics[scale=0.36]{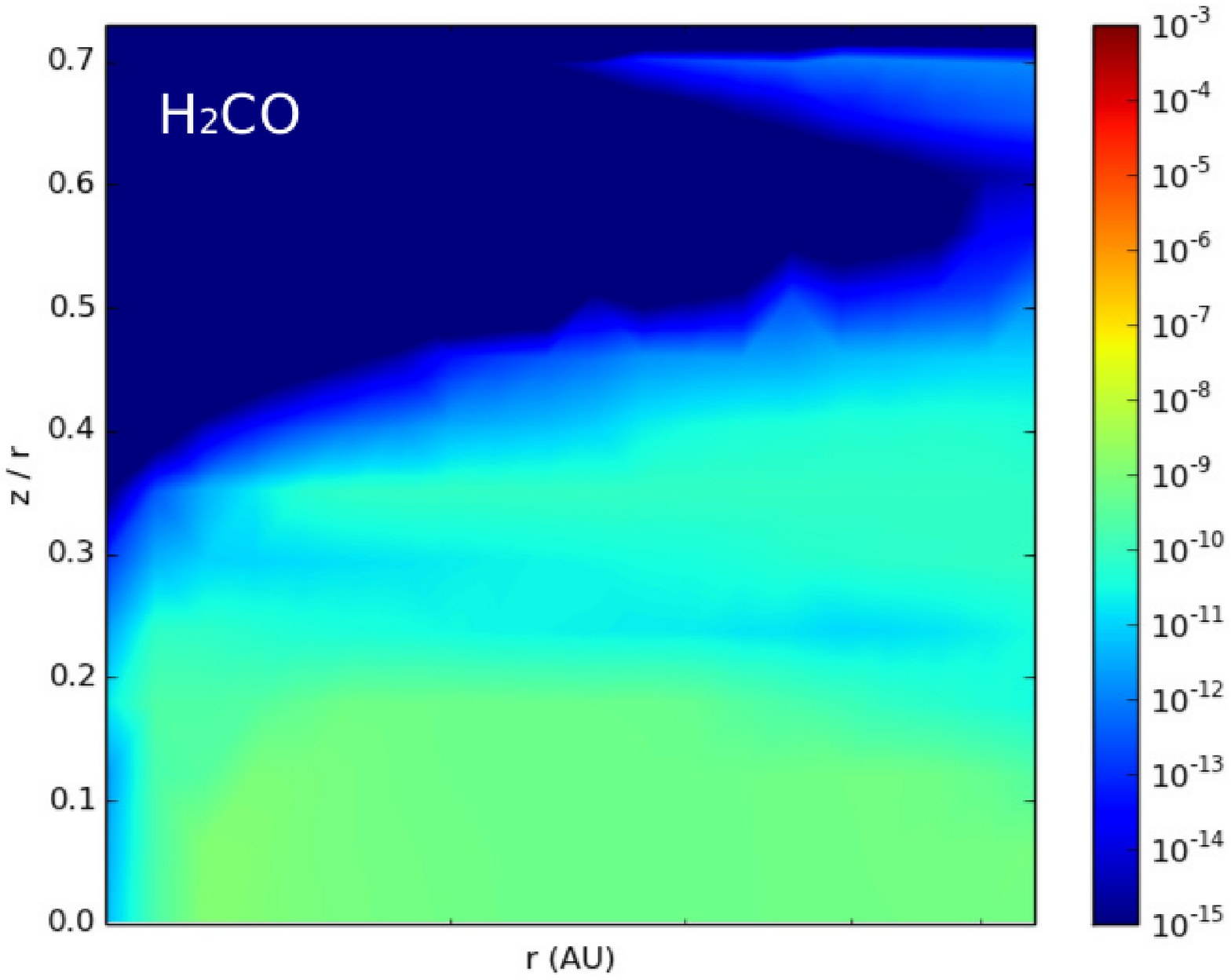}}
	\subfigure{\includegraphics[scale=0.36]{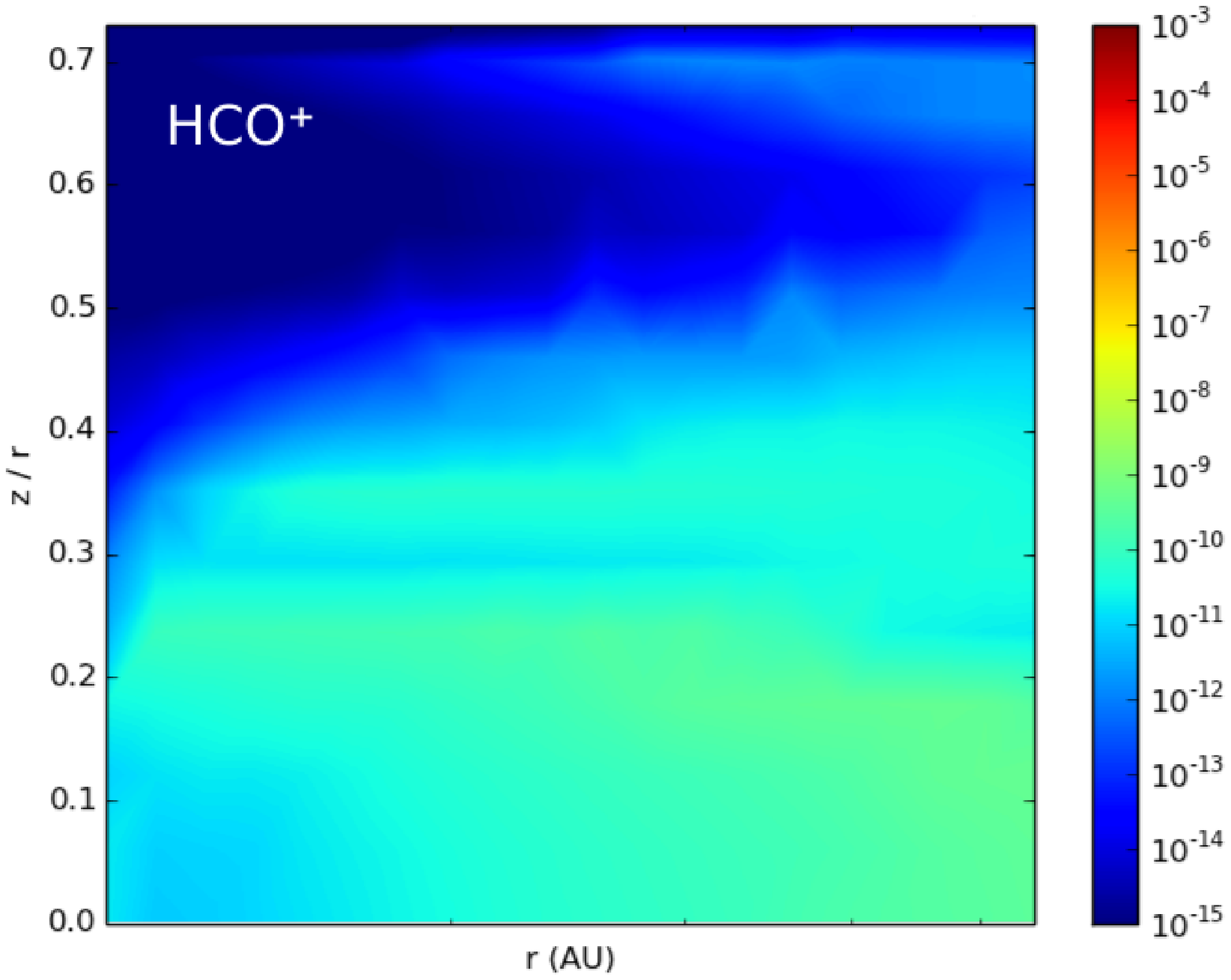}}
	\subfigure{\includegraphics[scale=0.36]{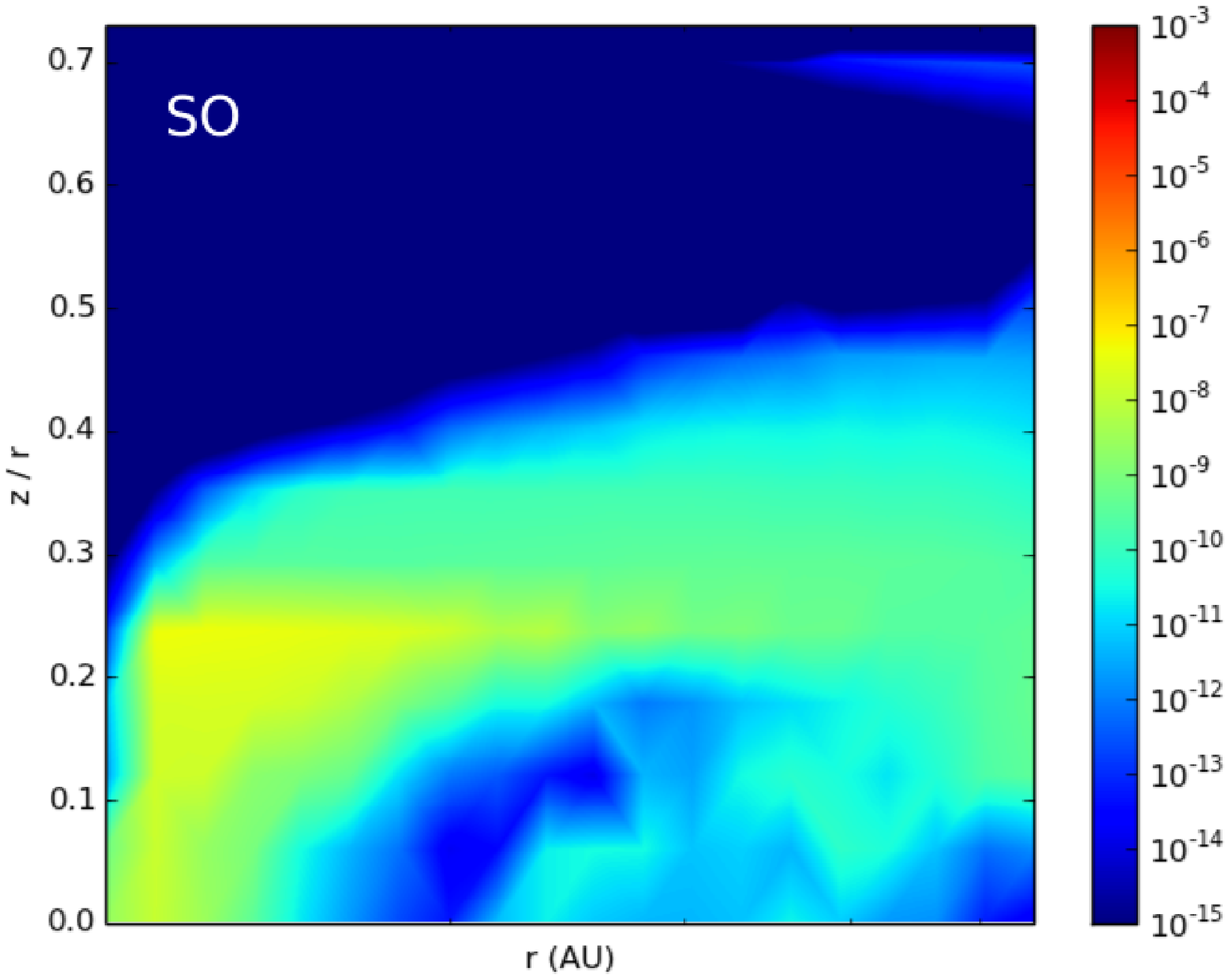}}
	\subfigure{\includegraphics[scale=0.36]{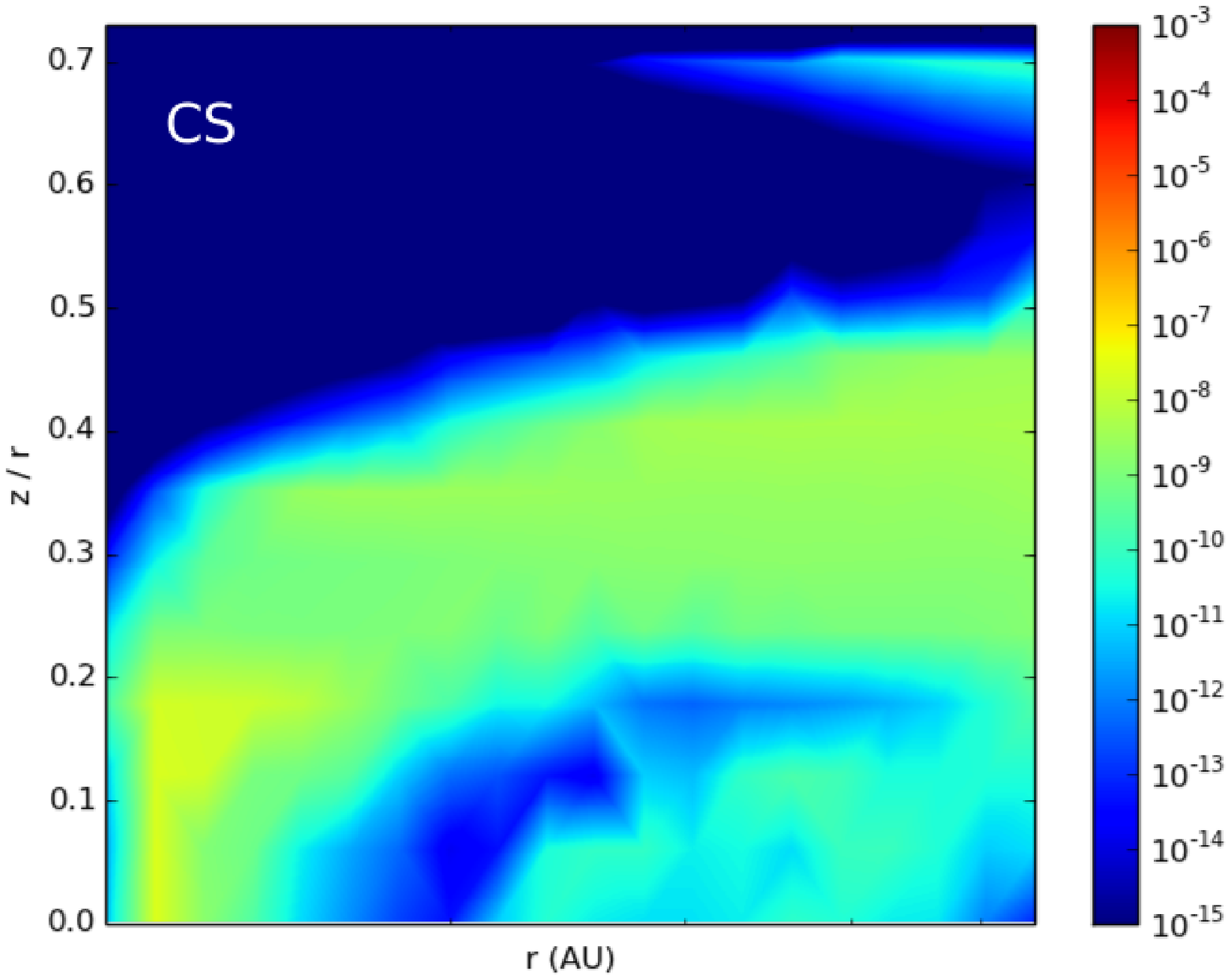}}
\caption{Molecule distribution of HCN, CN, H$_2$CO, HCO$^+$, SO, and CS as a function of
radius r and z/r (where z is the height over the midplane) for Model E.}
	\label{Fig:map_mole}%
\end{figure*}

\begin{figure*}
	\centering
	\subfigure{\includegraphics[scale=0.36]{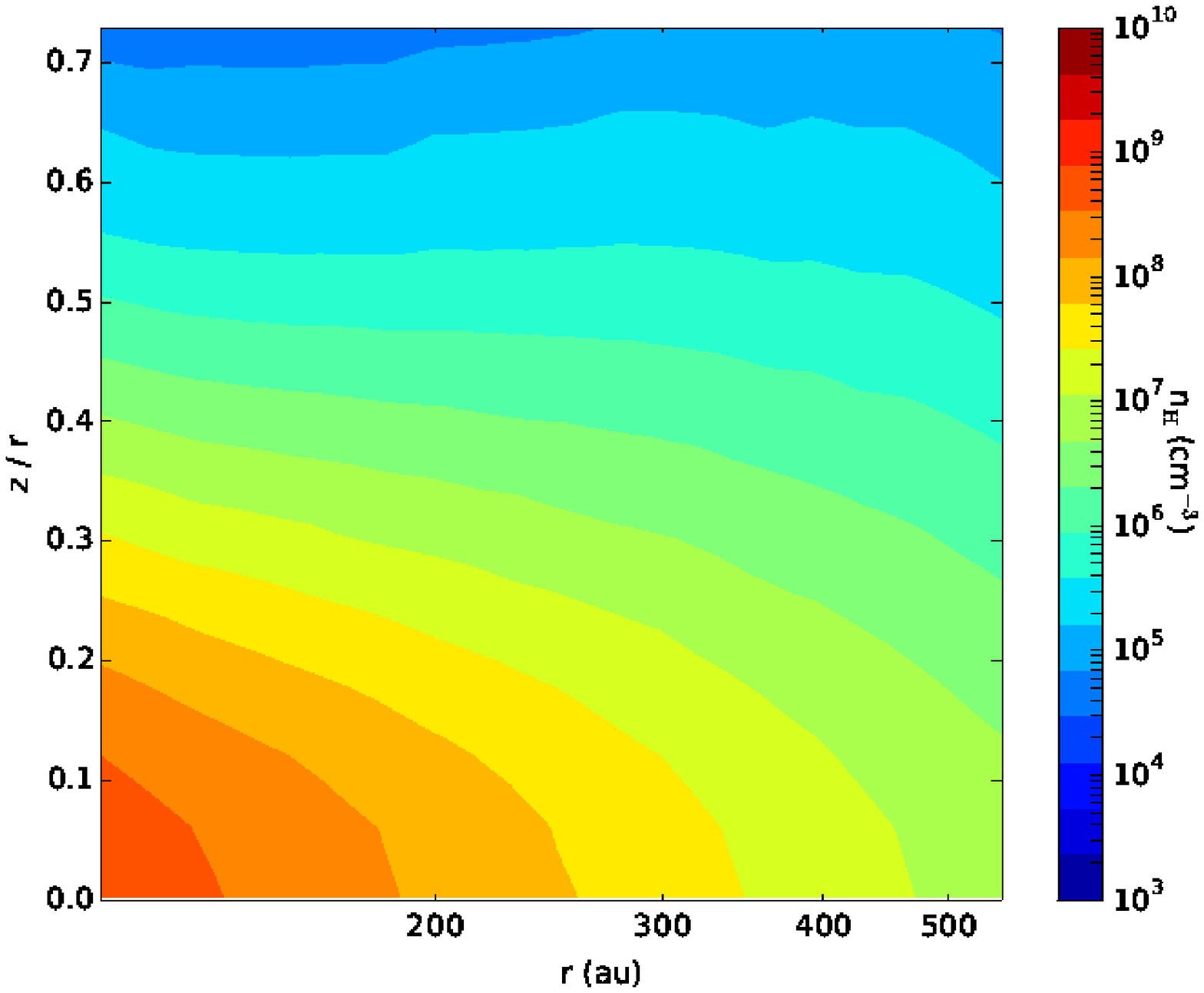}}
	\subfigure{\includegraphics[scale=0.36]{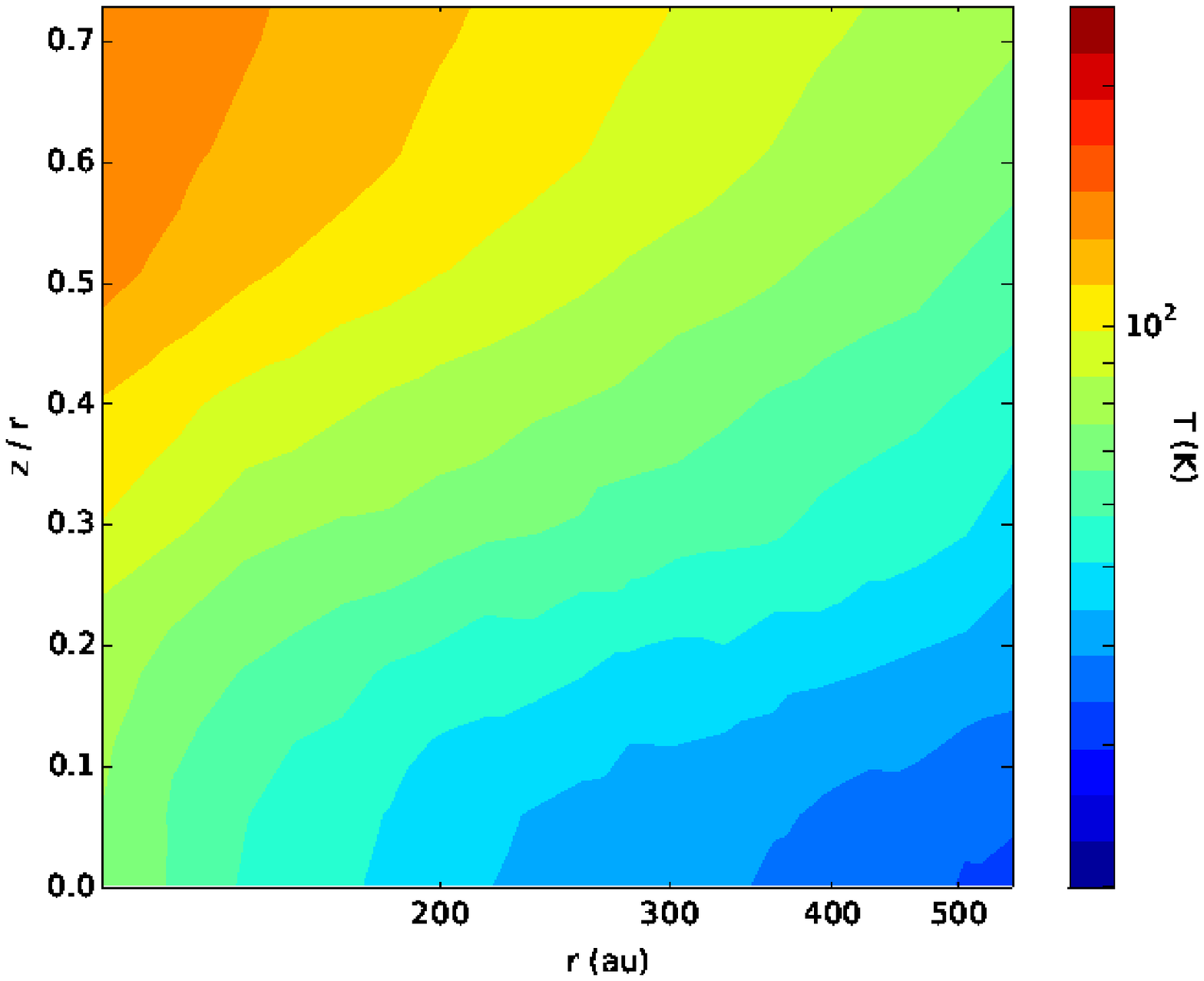}}
    \caption{Distribution of H nuclei volume density (\emph{left}) and temperature
(\emph{right}) as a function of radius r and z/r (where z is the height over the
midplane) in Model E.}
	\label{Fig:map_temp}%
\end{figure*}

\begin{figure*}[]
   \centering
	\subfigure{\includegraphics[scale=1]{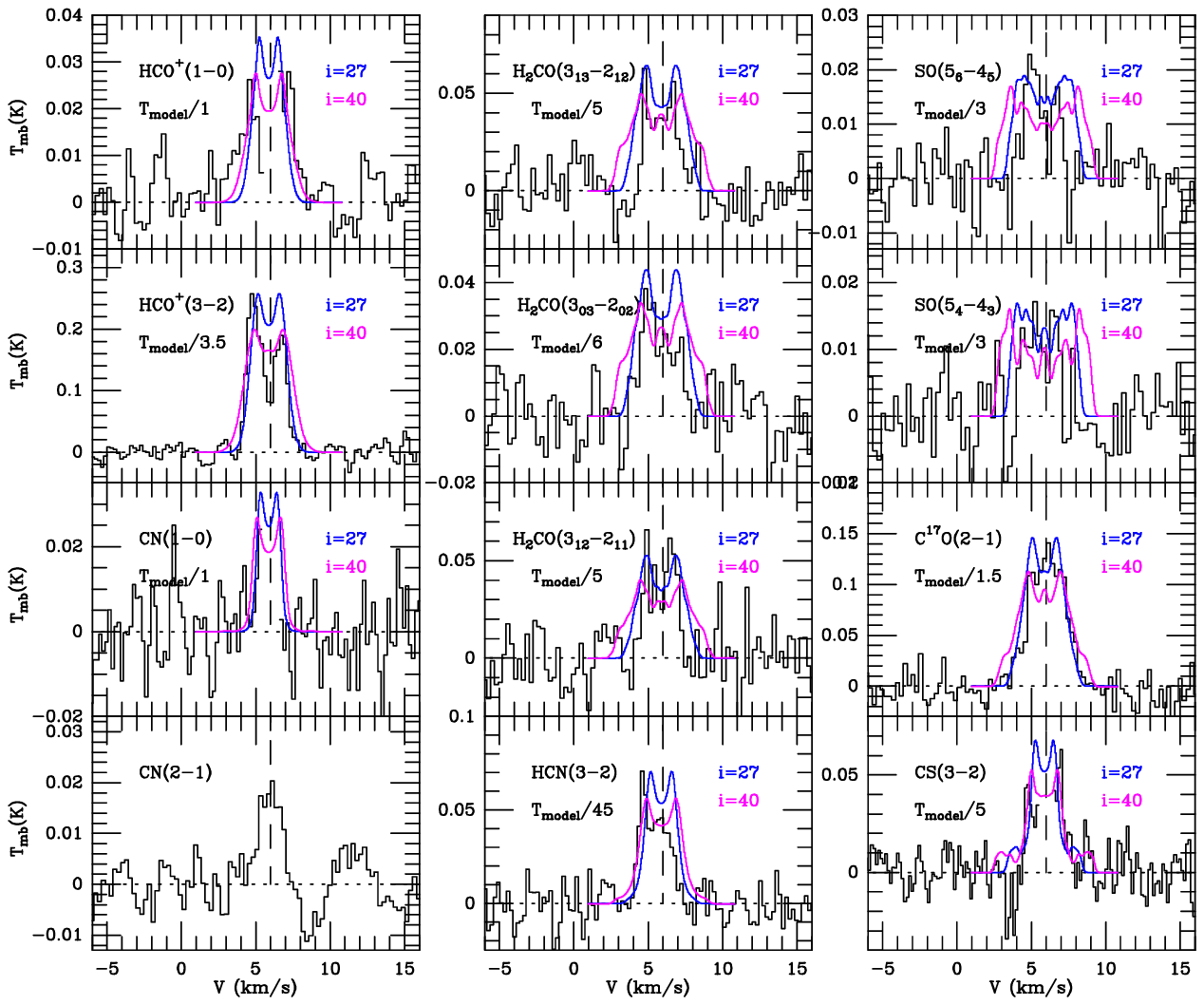}}
	\subfigure{\includegraphics[scale=1]{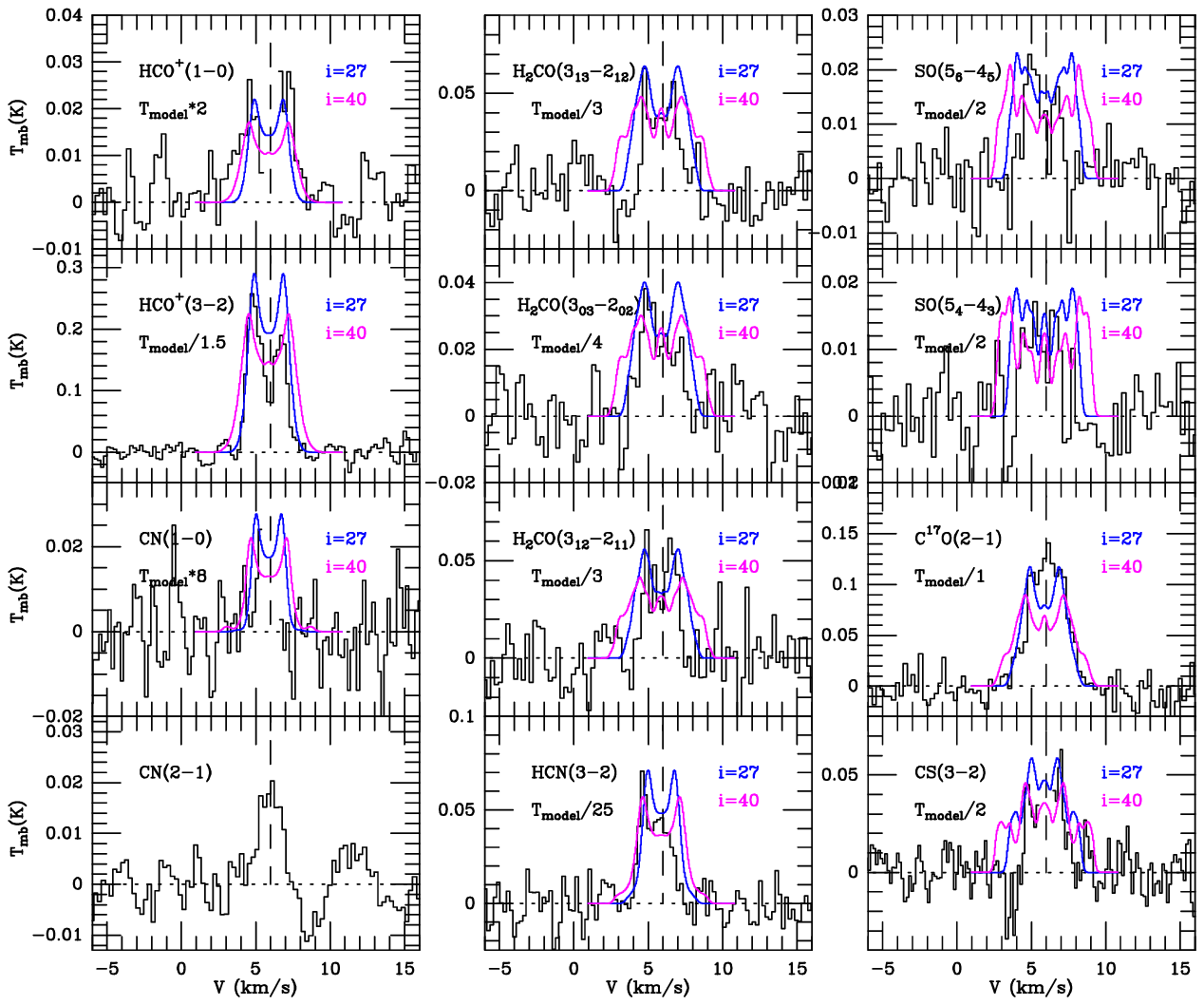}}
\caption{(\emph{Top}): Model A, (\emph{Bottom}): Model B. Comparison between modeled
spectra and those detected by 30m telescope toward
 AB Aur disk. In colors we
see the synthetic line profile obtain with our model. The \emph{blue} and
\emph{magenta} lines correspond to the same model with inclination angles for
the disk of 27$\degree$ and 40$\degree$ respectively. Vertical dash line
indicate the V$_{\rm{lsr}}$ = 5.9 \kms~ \citep{Duvert1986}.}	
\label{Fig:modelAB}
\end{figure*}

\begin{figure*}[]
   \centering
	\subfigure{\includegraphics[scale=1]{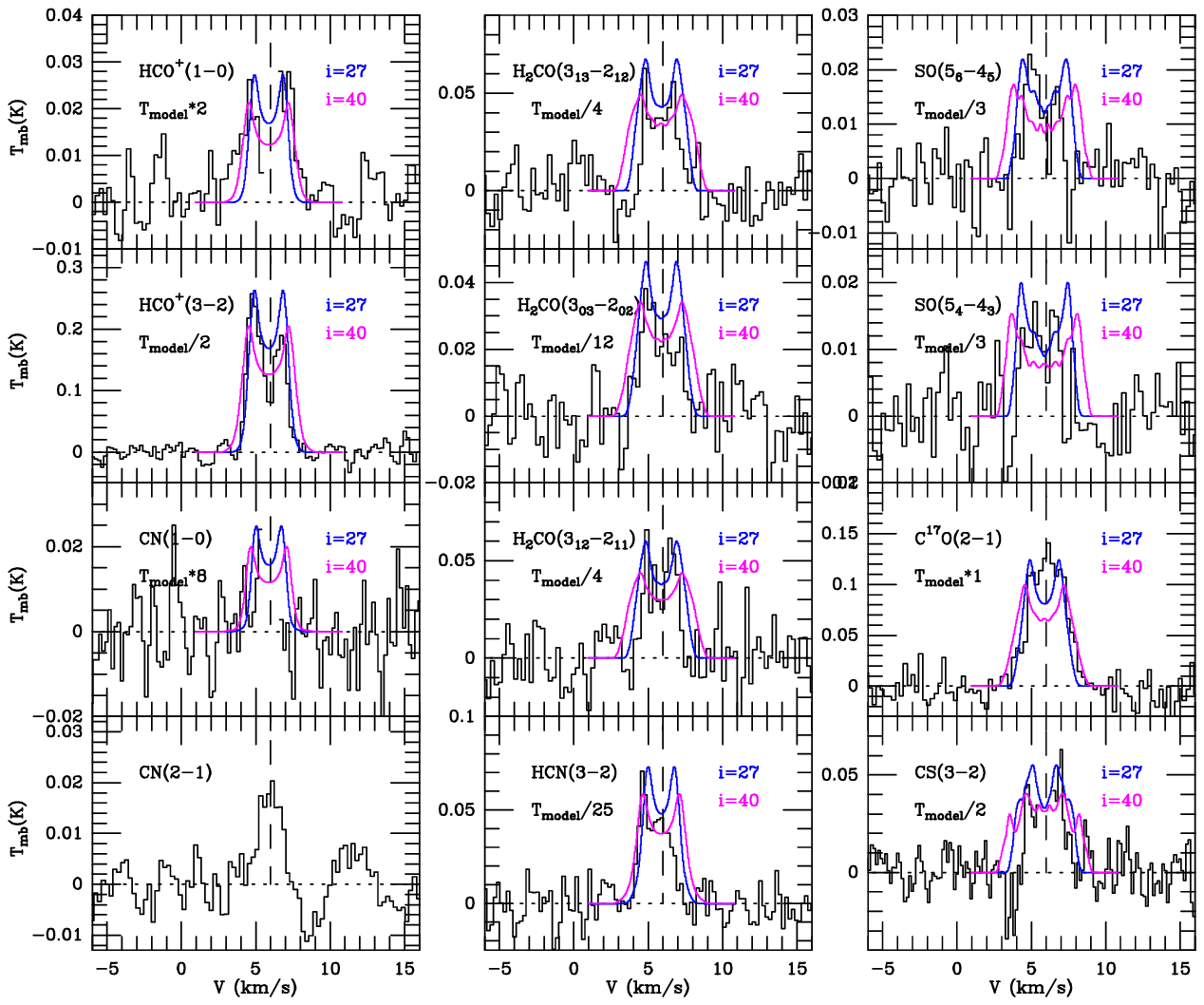}}
	\subfigure{\includegraphics[scale=1]{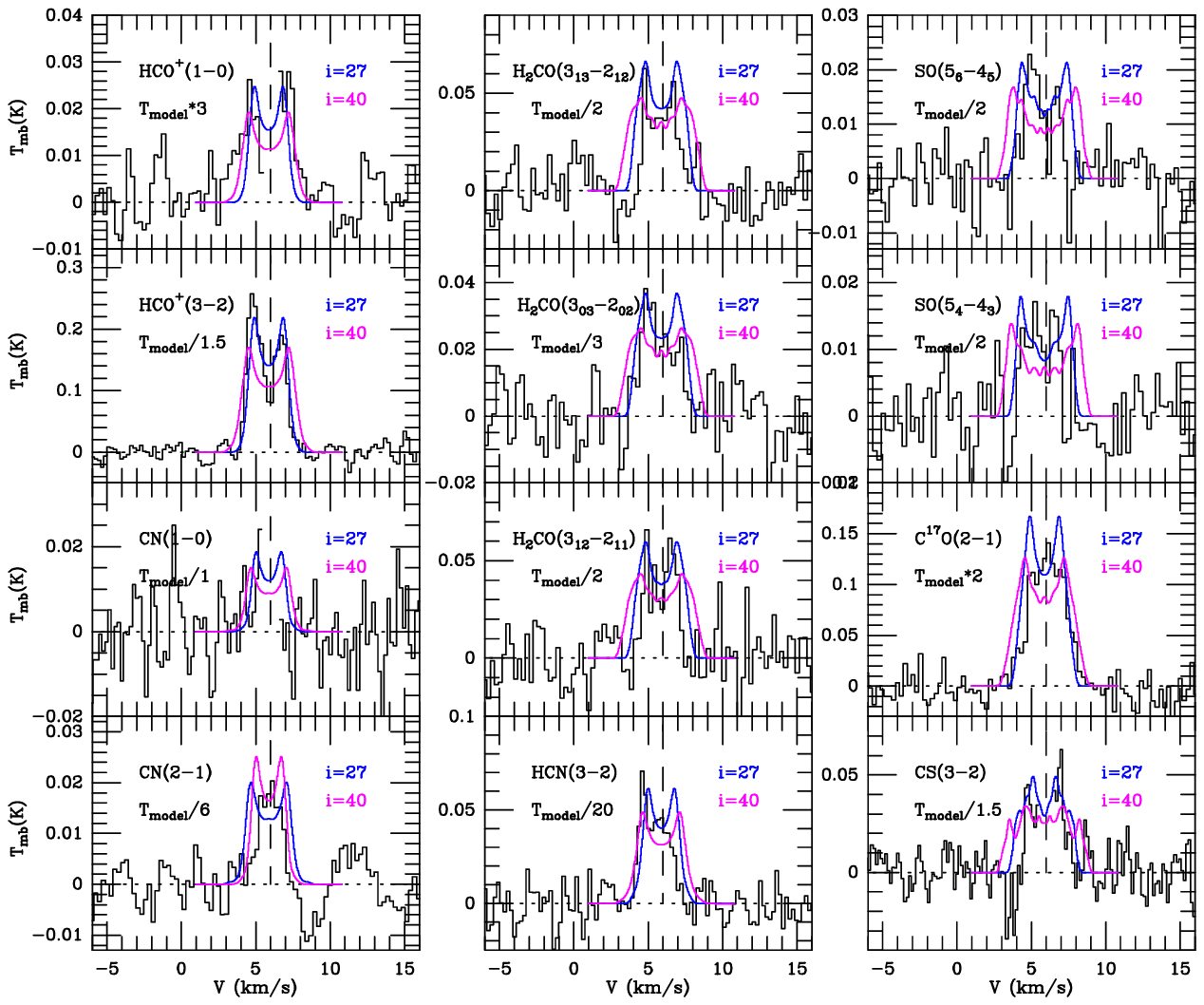}}
	\caption{(\emph{Top}): Model C, (\emph{bottom}): Model D.}
 \label{Fig:modelCD}
\end{figure*}

\end{appendix}

\end{document}